\documentclass[prd,nofootinbib,reprint,twocolumn,showpacs,superscriptaddress,floatfix]{revtex4-1}
\pdfoutput=1 
\usepackage{bm, color, xcolor} 
\usepackage{amssymb,amsfonts,slashed,amsthm,amsmath, graphicx, soul, empheq, cancel}
\usepackage{hyperref}
\usepackage[caption=false]{subfig}

\begin{document}

\newcommand{\Od}{ \mathcal{O} }
\newcommand{\vev}[1]{ \left\langle {#1} \right\rangle }
\newcommand{\bra}[1]{ \langle {#1} | }
\newcommand{\ket}[1]{ | {#1} \rangle }
\newcommand{\eV}{ \ {\rm eV} }
\newcommand{\KeV}{ \ {\rm keV} }
\newcommand{\MeV}{\  {\rm MeV} }
\newcommand{\GeV}{\  {\rm GeV} }
\newcommand{\TeV}{\  {\rm TeV} }
\newcommand{\1}{\mbox{1}\hspace{-0.25em}\mbox{l}}
\newcommand{\Red}[1]{{\color{red} {#1}}}

\newcommand{\lmk}{\left(}  
\newcommand{\rmk}{\right)}
\newcommand{\lkk}{\left[}  
\newcommand{\rkk}{\right]}
\newcommand{\lhk}{\left \{ }  
\newcommand{\rhk}{\right \} }
\newcommand{\del}{\partial}  
\newcommand{\la}{\left\langle} 
\newcommand{\ra}{\right\rangle}
\newcommand{\half}{\frac{1}{2}}

\newcommand{\bea}{\begin{array}}
\newcommand{\eea}{\end{array}}
\newcommand{\beq}{\begin{eqnarray}}
\newcommand{\eeq}{\end{eqnarray}}
\newcommand{\eq}[1]{Eq.~(\ref{#1})}

\newcommand{\dd}{\mathrm{d}}
\newcommand{\Mpl}{M_{\rm Pl}}
\newcommand{\mg}{m_{3/2}}
\newcommand{\abs}[1]{\left\vert {#1} \right\vert}
\newcommand{\mphi}{m_{\phi}}
\newcommand{\Hz}{\ {\rm Hz}}
\newcommand{\for}{\quad \text{for }}
\newcommand{\Min}{\text{Min}}
\newcommand{\Max}{\text{Max}}
\newcommand{\Kahler}{K\"{a}hler }
\newcommand{\cphi}{\varphi}
\newcommand{\Tr}{\text{Tr}}
\newcommand{\diag}{{\rm diag}}

\newcommand{\SUf}{SU(3)_{\rm f}}
\newcommand{\Upq}{U(1)_{\rm PQ}}
\newcommand{\Zpq}{Z^{\rm PQ}_3}
\newcommand{\Cpq}{C_{\rm PQ}}
\newcommand{\ubar}{u^c}
\newcommand{\dbar}{d^c}
\newcommand{\ebar}{e^c}
\newcommand{\nubar}{\nu^c}
\newcommand{\Ndw}{N_{\rm DW}}
\newcommand{\Fpq}{F_{\rm PQ}}
\newcommand{\fpq}{v_{\rm PQ}}
\newcommand{\Br}{{\rm Br}}
\newcommand{\Lag}{\mathcal{L}}
\newcommand{\Lqcd}{\Lambda_{\rm QCD}}

\newcommand{\ji}{j_{\rm inf}} 
\newcommand{\jb}{j_{B-L}} 
\newcommand{\M}{M} 
\newcommand{\im}{{\rm Im} }
\newcommand{\re}{{\rm Re} }

\newcommand{\tdelta}{\tilde{\delta}}
\newcommand{\tsigma}{\tilde{\sigma}}
\newcommand{\tPhi}{\tilde{\Phi}}

\def\lrf#1#2{ \left(\frac{#1}{#2}\right)}
\def\lrfp#1#2#3{ \left(\frac{#1}{#2} \right)^{#3}}
\def\lrp#1#2{\left( #1 \right)^{#2}}
\def\REF#1{Ref.~\cite{#1}}
\def\SEC#1{Sec.~\ref{#1}}
\def\FIG#1{Fig.~\ref{#1}}
\def\EQ#1{Eq.~(\ref{#1})}
\def\EQS#1{Eqs.~(\ref{#1})}
\def\TEV#1{10^{#1}{\rm\,TeV}}
\def\GEV#1{10^{#1}{\rm\,GeV}}
\def\MEV#1{10^{#1}{\rm\,MeV}}
\def\KEV#1{10^{#1}{\rm\,keV}}
\def\blue#1{\textcolor{blue}{#1}}
\def\red#1{\textcolor{blue}{#1}}

\newcommand{\eff}{\Delta N_{\rm eff}}
\newcommand{\neff}{\Delta N_{\rm eff}}
\newcommand{\cc}{\Omega_\Lambda}
\newcommand{\Mpc}{\ {\rm Mpc}}
\newcommand{\Msolar}{M_\odot}

\def\sn#1{\textcolor{red}{#1}}
\def\SN#1{\textcolor{red}{[{\bf SN:} #1]}}

\def\jx#1{\textcolor{blue}{#1}}
\def\JX#1{\textcolor{blue}{[{\bf JX:} #1]}}

\def\yw#1{\textcolor{teal}{(Note updated \today) #1}}
\def\YW#1{\textcolor{teal}{[{\bf YW (note updated \today):} #1]}}

\def\NK#1{\textcolor{magenta}{[{\bf NK:} #1]}}

\title{Toward Simulations of Cold Baryogenesis via Supercooled Phase Transitions}

\author{Naoya Kitajima}
\email{naoya.kitajima.c2@tohoku.ac.jp}
\affiliation{Frontier Research Institute for Interdisciplinary Sciences, Tohoku University, \\
6-3 Azaaoba, Aramaki, Aoba-ku, Sendai 980-8578, Japan}
\affiliation{Department of Physics, Tohoku University, \\
6-3 Azaaoba, Aramaki, Aoba-ku, Sendai 980-8578, Japan}
\author{Shota Nakagawa}
\email{shota.nakagawa@sjtu.edu.cn}
\author{Yuichiro Nakai}
\email{ynakai@sjtu.edu.cn}
\author{Yaoduo Wang}
\email{yaoduowang@sjtu.edu.cn}
\author{Junxuan Xu}
\email{jxxu0108@sjtu.edu.cn}
\affiliation{Tsung-Dao Lee Institute, Shanghai Jiao Tong University, \\
No.~1 Lisuo Road, Pudong New Area, Shanghai 201210, China}
\affiliation{School of Physics and Astronomy, Shanghai Jiao Tong University, \\
800 Dongchuan Road, Shanghai 200240, China}

\begin{abstract}
Supercooled first-order phase transitions provide an attractive framework for generating observable gravitational waves (GWs), but the accompanying entropy injection dilutes any pre-existing baryon asymmetry, motivating baryogenesis during or after the phase transition. Cold baryogenesis offers a promising mechanism in which bubble collisions generate Higgs winding configurations and the Chern-Simons number through out-of-equilibrium dynamics. In the present study, we perform three-dimensional lattice simulations of a two-field model where the scalar field responsible for the phase transition is separated from the Higgs sector. Using realistic initial conditions obtained from bounce solutions, we follow the bubble expansion and collisions and study the generation of Chern-Simons number. It is found that efficient Chern-Simons production persists even for strongly non-degenerate potentials with large latent heat, which are favorable for enhancing GW signals.
We also estimate the resulting baryon asymmetry using the simulated Chern-Simons dynamics and an effective CP-violating operator. Our results demonstrate that cold baryogenesis remains viable in the strongly supercooled regime, providing an important step toward establishing baryogenesis from cosmological first-order phase transitions.
\end{abstract}

\maketitle


\section{Introduction
\label{sec:introduction}}

First-order phase transitions in the early Universe have recently gained considerable attention as viable sources of a stochastic gravitational wave (GW) background. In particular, {\it a supercooled phase transition}, in which the Universe remains trapped in a false vacuum until far below the critical temperature, can maximize the GW amplitude. The produced GW spectrum is able to provide a promising interpretation of the signal observed by pulsar timing array (PTA) experiments
\cite{NANOGrav:2023gor,NANOGrav:2023hvm,EPTA:2023fyk,Reardon:2023gzh,Xu:2023wog} (see Refs.~\cite{Nakai:2020oit,Fujikura:2023lkn,Madge:2023dxc,Megias:2023kiy,Salvio:2023ynn,Salvio:2023blb,Gouttenoire:2023bqy,Addazi:2023jvg,Li:2023bxy,Ghosh:2023aum,Jiang:2023qbm,Wang:2023bbc,Li:2025nja,Fujikura:2025iam}). At higher frequencies, supercooled phase transitions are also expected to generate GW signals in the milli-hertz range accessible to space-based interferometers such as LISA~\cite{LISA:2017pwj}, DECIGO~\cite{Kawamura:2020pcg}, BBO~\cite{Harry:2006fi}, TianQin~\cite{TianQin:2015yph,TianQin:2020hid} and Taiji~\cite{Hu:2017mde}, providing a unique opportunity to probe physics beyond the Standard Model (SM).

A critical implication of a supercooled phase transition is that the vacuum-energy-dominated stage preceding the phase transition generically dilutes baryon asymmetry generated at earlier times. During the supercooling phase, the rapid expansion of the Universe exponentially suppresses pre-existing number densities, rendering conventional high-scale baryogenesis mechanisms ineffective unless the transition takes place promptly. Consequently, a supercooled phase transition naturally motivates a baryogenesis scenario that operates at or after the transition. In particular, it is a fascinating possibility that the out-of-equilibrium dynamics of bubble nucleation, expansion and collision in a supercooled phase transition plays a central role in baryogenesis. The same dynamics responsible for enhanced GW signals may also account for the origin of matter-antimatter asymmetry in the Universe.

One potential realization of baryogenesis via a supercooled phase transition is based on the interplay between Higgs winding dynamics and chiral anomaly
\cite{Konstandin:2011ds,Fujikura:2024jto}, referred to as {\it cold baryogenesis} which was originally proposed in the context of low-scale hybrid inflation
\cite{Krauss:1999ng,Garcia-Bellido:1999xos}. The expected physics picture is then given by the analogy with post-inflationary preheating. During a supercooled phase transition, collisions of true-vacuum bubbles drive a scalar field responsible for the phase transition, denoted as $\sigma$, into large oscillations around its vacuum expectation value (VEV). In the nearly empty Universe created by supercooling, these oscillations induce tachyonic instability for an $SU(2)$ doublet Higgs field $\Phi$ through its coupling to $\sigma$, triggering spontaneous breaking of an $SU(2)$ gauge symmetry. Similar to inflationary preheating, this process may abundantly produce long-wavelength Higgs modes. Due to bubble collision dynamics, the produced Higgs field may develop spatially inhomogeneous $SU(2)$ orientations, resulting in the formation of Higgs field configurations carrying nonzero winding number. When coupled to the gauge field, these configurations subsequently relax toward the vacuum either by unwinding the Higgs field itself or by inducing a change in the Chern-Simons number of the gauge field. Since such a variation of the Chern-Simons number corresponds to fermion number violation through chiral anomaly, a nonzero asymmetry can be generated provided C and CP symmetries are violated. Configurations with positive and negative differences between the Chern-Simons number and the Higgs winding number evolve asymmetrically: those with one sign preferentially relax via gauge-field transitions, while those with the opposite sign tend to unwind through Higgs-field dynamics. This small bias in the relaxation channels ultimately produces a net asymmetry.

Despite the reasonable analogy with the case of post-inflationary preheating, it remains a non-trivial issue to establish that the  mechanism of cold baryogenesis operates efficiently on a supercooled phase transition rather than an inflationary background. Due to the highly nonlinear dynamics, a definitive assessment requires a fully non-perturbative lattice simulation that directly tracks the real-time evolution of a scalar field responsible for the phase transition $\sigma$ and the Higgs-gauge system coupled to $\sigma$. The authors of Ref.~\cite{Bhusal:2025lvm} have recently initiated to perform a lattice simulation in this direction where they considered a first-order phase transition of the Higgs field responsible for cold baryogenesis.
In this simplified setup, they defined a degeneracy parameter characterizing the Higgs potential shape,
$\epsilon \equiv \frac{V_{\rm b}}{V_{\rm b} + \Delta V}$ where $V_{\rm b}$ denotes the barrier height and $\Delta V$
is the energy difference between the false and true vacua,
and concluded that a small $\epsilon$, which is favored for the production of a large GW amplitude,
is disfavored for cold baryogenesis.
This is because when the false and true vacua are degenerate, colliding walls can penetrate one another, allowing the interaction region to persist over an extended space volume and enabling the generation of the Chern-Simons number well beyond the initial collision point. By contrast, in the case with non-degenerate vacua, bubble collisions are confined to narrow regions near the bubble surfaces. In the present paper, we perform a lattice simulation for the full setup of the scalar field $\sigma$ coupled to the Higgs-gauge system, and show that the Chern-Simons number can be abundantly produced for the case with non-degenerate vacua in the $\sigma$ potential.

The phase transition duration $\beta^{-1}$ is a key parameter to quantify how rapidly the transition completes in time,
and a supercooled phase transition is characterized by a small value of $\beta/H = \mathcal{O} (1-10)$ where $H$ denotes the Hubble parameter. 
One difficulty in performing lattice simulations for a supercooled phase transition is that the bubble radius at collision $R_*$ is large and comparable to $H^{-1}$, while the wall thickness becomes very small due to Lorentz contraction, requiring a large hierarchy of spatial scales to be resolved with a huge computational cost. By taking a sufficiently large phase-transition VEV $v_\sigma$, however, the bubble size can be reduced while keeping $\beta/H$ relatively small, making the supercooled regime accessible to our lattice simulations.
We still study the dependence of our results on the Lorentz factor $\gamma_*$ and use it to extrapolate toward the regime of smaller $v_\sigma$ and larger $\gamma_*$.

The rest of the paper is organized as follows.
In section~\ref{vacuum transition}, we introduce our setup and discuss the nucleation of bubbles in our first-order phase transition.
Section~\ref{sec:dynamics} investigates the evolution of the system after the bubble nucleation,
and describe the amount of fermion number asymmetry produced by the present system.
Then, in section~\ref{sec:simulation}, we present our lattice simulation setup and show the results.
Section~\ref{sec:Discussion} is devoted to conclusions and discussions.
Lattice simulation details are summarized in appendices.

\section{Vacuum transition}
\label{vacuum transition}

In this section, we introduce our two-field model and discuss the vacuum transition. We then study the bubble nucleation based on the bounce solution and identify the parameter region relevant for our supercooled phase transition.

\subsection{Setup
\label{sec:Setup}}

We consider an $SU(2)$ gauge theory containing a doublet Higgs field
$\Phi\equiv (\Phi_u,\Phi_d)^T$ coupled to a real singlet scalar field $\sigma$.
The Lagrangian is given by
\begin{align}
\Lag = -\frac{1}{2}\Tr[F_{\mu\nu}F^{\mu\nu}] - |D_\mu \Phi|^2 - \frac{1}{2}(\del_\mu\sigma)^2 -V(\Phi, \sigma) \ .
\end{align}
Here $A_\mu = A_\mu^a \tau^a/2$ denotes the $SU(2)$ gauge field,
where $\tau^a$ ($a=1,2,3$) are the Pauli matrices. The corresponding
field-strength tensor and the covariant derivative acting on the Higgs
doublet are defined as
\begin{align}
F_{\mu\nu} &\equiv \partial_\mu A_\nu-\partial_\nu A_\mu
-ig[A_\mu,A_\nu] \ , \\[1ex]
D_\mu &\equiv \partial_\mu-igA_\mu \ ,
\end{align}
where $g$ denotes the $SU(2)$ gauge coupling.
We consider the following zero-temperature potential,
\begin{align}
V(\Phi, \sigma) = V_{\rm eff}(\sigma) + \frac{1}{4} \lambda_\Phi \left[|\Phi|^2 -\frac{v_\Phi^2}{2} \lmk\frac{\sigma}{v_\sigma}\rmk^2\right]^2,
\label{potential}
\end{align}
with an effective potential for $\sigma$,\footnote{This potential form is motivated by an extra-dimensional model with the Goldberger-Wise mechanism \cite{Randall:1999ee,Goldberger:1999uk}.
Although there is a limited parameter space for a sufficient nucleation rate which was well-studied in Ref.~\cite{vonHarling:2017yew} (see also Refs.~\cite{Creminelli:2001th,Baratella:2018pxi,Fujikura:2019oyi}), we use it as a toy model of a supercooled phase transition.}
\begin{align}
V_{\rm eff}(\sigma) = \lambda_\sigma\sigma^4\left[(4+2a)\lmk 1-x\lmk\frac{|\sigma|}{v_\sigma}\rmk^a\rmk^2 -a +b\right].
\label{Veff}
\end{align}
Here $\lambda_\Phi,\lambda_\sigma~(>0)$, $a~(\neq0)$, and $b$ are dimensionless constants, and $v_\Phi$ and $v_\sigma$ represent the VEVs of $\Phi$ and $\sigma$ at the true minimum, respectively.
For simplicity, we take $\lambda_\Phi=\lambda_\sigma=1$ and $v_\Phi = 0.8 v_\sigma$ throughout the present paper.
The magnitude of $a$ indicates the degree of violation of scale invariance, which can be understood from the fact that $V_{\rm eff}\propto\sigma^4$ in the limit of $a\rightarrow0$.
When we take the value $v_\sigma$ as an independent parameter, a parameter $x$ is parametrized by $a,b$ as
\begin{align}
x = \lmk1+\frac{a}{2}\rmk^{-1}\lmk1+\frac{a}{4}+\frac{{\rm sgn}(a)}{2}\sqrt{a+\frac{a^2}{4}-b}\rmk.
\label{xpara}
\end{align}
Solving the stationary condition, we find the local maximum at $\sigma_{b}=(y/x)^{1/a}v_\sigma$ with 
\begin{align}
y = \lmk1+\frac{a}{2}\rmk^{-1}\lmk1+\frac{a}{4}-\frac{{\rm sgn}(a)}{2}\sqrt{a+\frac{a^2}{4}-b}\rmk.
\label{ypara}
\end{align}

\begin{figure}[t!]
\centering
\includegraphics[width=8.5cm]{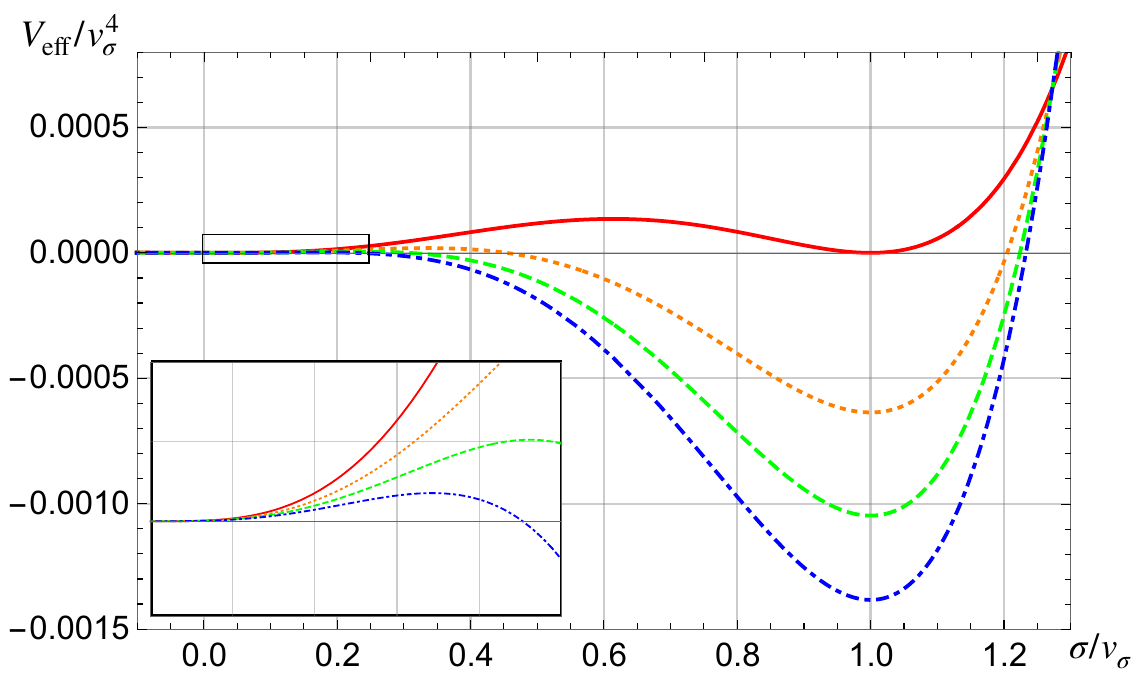}
\caption{
The shape of the potential (\ref{Veff}) normalized by $v_\sigma^4$ as a function of $\sigma/v_\sigma$.
Here we set $a=0.1$.
The red, orange dotted, green dashed, and blue dot-dashed lines respectively correspond to $b/a=1,0.9,0.8,0.7$ ($\epsilon \simeq 1, 0.029, 0.0049, 0.0013$).
The small barrier can be seen in the mini-window inside the figure.
}
\label{fig:potential}
\end{figure}

When there is a barrier in the potential $V_{\rm eff}$, the system undergoes a first-order phase transition, which induces the spontaneous breaking of $SU(2)$ due to mixing between $\Phi$ and $\sigma$ in Eq.~\eqref{potential}.
To have the barrier, we take $a,b$ as
\begin{align}
-(4+a) < b < a \ .
\label{b_range}
\end{align}
One can see from Eq.~\eqref{xpara} or (\ref{ypara}) that we have a range $-(4+a) < b < a+a^2/4$ to obtain the two local minima.
For $b>a$, the minimum at $\sigma=v_\sigma$ is higher in energy than
that at $\sigma=0$ and therefore ceases to be the true vacuum. This
sets the upper bound $b<a$.
We can define the degeneracy parameter as \cite{Jinno:2019bxw}
\begin{align}
\epsilon &\equiv \frac{V_{\rm b}}{\Delta V + V_{\rm b}} \in(0,1) \ ,
\label{degeneracy}
\end{align}
where we define the barrier height and the potential energy difference between the false and true minima as $V_{\rm b}\equiv V_{\rm eff}(\sigma_{b})-V_{\rm eff}(0)$ and $\Delta V \equiv V_{\rm eff}(0)-V_{\rm eff}(v_\sigma)$, respectively.
The larger the value of $\epsilon$ is, the more degenerate the local minima, and $\epsilon=1~(b=a)$ shows the exact degeneracy (see \FIG{fig:potential}).

Finite-temperature corrections to the effective potential can induce thermal tunneling. We focus, however, on a strongly supercooled phase transition, in which bubble nucleation occurs at a temperature much lower than the critical temperature. In this regime, finite-temperature corrections are expected to have only a minor effect on the bounce action. We therefore assume that bubble nucleation is dominated by quantum tunneling governed by the zero-temperature potential.

\subsection{Bubble nucleation 
\label{sec:bubble_nucleation}}

The first-order phase transition proceeds through the nucleation of bubbles characterized by nontrivial configurations of both fields, $\sigma_c$ and $\Phi_c=U(\theta_i)(0,h_c/\sqrt{2})$. 
Here $U(\theta_i)=e^{i\tau_i\theta_i}$   with $\theta_i$ the orientation in the $SU(2)$ space, which is randomly assigned to each bubble.
The bubble nucleation rate can be estimated by using a semi-classical approximation \cite{Coleman:1985rnk}.
In the absence of finite-temperature effects, the bubble nucleation is described by the $O(4)$-symmetric bounce equations,
\begin{equation}
\begin{aligned}
&\frac{d^2h_c}{d\rho_E^2} + \frac{3}{\rho_E}\frac{dh_c}{d\rho_E}=\frac{\del V(h_c,\sigma_c)}{\del h_c} \ ,\\[1ex]
&\frac{d^2\sigma_c}{d\rho_E^2} + \frac{3}{\rho_E}\frac{d\sigma_c}{d\rho_E}=\frac{d V(h_c,\sigma_c)}{d \sigma_c} \ ,
\label{bounce_eq}
\end{aligned}
\end{equation}
with the boundary conditions, $\sigma_c(\infty)=h_c(\infty)=0$ and $\sigma_c'(0)=h_c'(0)=0$.
Here we define $\rho_E\equiv\sqrt{r^2+\tau^2}$ with $r$ the radial coordinate in the three-dimensional space and $\tau$ the Euclidean time.

\begin{figure}[t!]
\centering
\includegraphics[width=8cm]{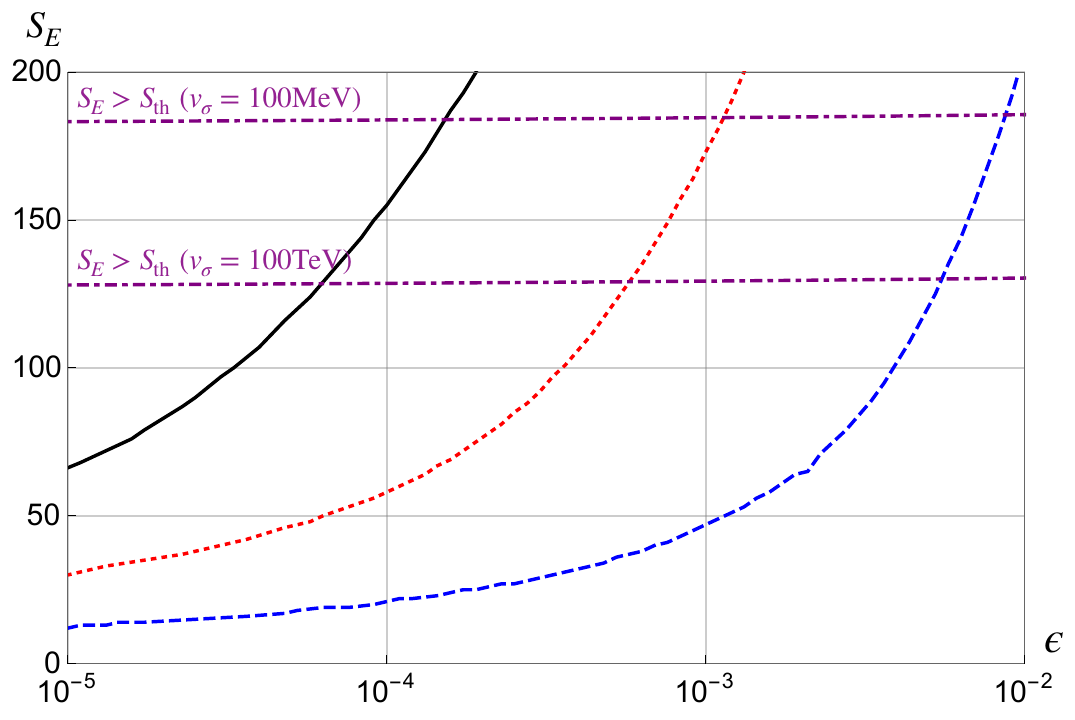}
\caption{
The bounce action as a function of $\epsilon$. The black, red dotted and blue dashed lines correspond to the cases with $a=0.15,0.2,0.3$, respectively.
Below the purple dot-dashed lines, the nucleation is efficient.
}
\label{fig:bounce}
\end{figure}

As the nucleation rate is given by $\Gamma\sim v_\sigma^4 e^{-S_E}$ with $S_E$ the Euclidean action, the bounce action should be small enough. 
When $\Gamma \sim H^4$, where $H$ is the Hubble parameter, the nucleation becomes effective, and the temperature at this time is defined as the nucleation temperature.
Although we cannot estimate the nucleation temperature because the thermal effect is ignored, we can use the bounce action at the zero-temperature to obtain the necessary condition for the nucleation to be completed.
Considering the Universe is dominated by the vacuum energy $(\sim v_\sigma^4)$ of $\sigma$, the necessary condition is given by $\Gamma/H^4\gtrsim1$, or \begin{align}
S_E\lesssim S_{\rm th}=2\ln\left[\frac{3\Mpl^2v_\sigma^2}{\Delta V}\right].
\label{Sth}
\end{align}
\FIG{fig:bounce} describes the bounce action as a function of $\epsilon$ for $a=0.15,0.2,0.3$, denoted by the black, red dotted, and blue dashed lines, respectively.
Here, we have calculated the bounce action by using FindBounce \cite{Guada:2018jek,Guada:2020xnz}.
For reference, the threshold values in Eq.~(\ref{Sth}) are indicated by the horizontal purple dot-dashed lines for $v_\sigma = 100 \MeV$ and $100 \TeV$.

\begin{figure}[t!]
\centering
\includegraphics[width=8.5cm]{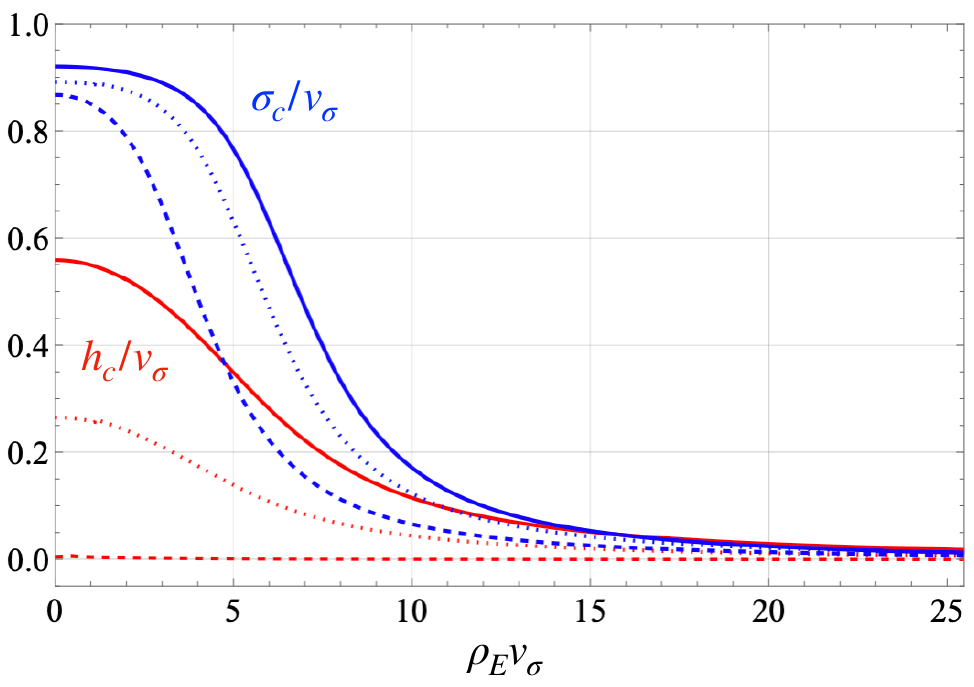}
\caption{
The field configurations of $\sigma_c$ (blue, the upper three lines) and $h_c$ (red, the lower three lines) at the nucleation for $a=0.3$.
The solid, dotted and dashed lines correspond to the cases with $\epsilon=10^{-2}$, $5\times10^{-3}$ and $10^{-3}$, respectively.
}
\label{fig:config}
\end{figure}

At the center of the bounce configuration ($\tau=0$), the bounce solution is analytically continuated with the Minkowski spacetime.
Thus, the bounce configuration provides the bubble profile at the nucleation which will be used as the initial condition when we perform lattice simulations in \SEC{sec:simulation}.
\FIG{fig:config} shows the bounce configurations for $a=0.3$ and $\epsilon=10^{-2}$ (solid line), $5\times10^{-3}$ (dotted line) and $10^{-3}$ (dashed line).
The exit point decreases with $\epsilon$.
For smaller $\epsilon$, the Higgs configuration becomes negligibly small, because the true vacuum is almost along the line of $h_c=0$, as the exit point of $\sigma$ gets closer to the barrier.
In this case, the nucleation proceeds dominantly with the field $\sigma$, and the Higgs field $h_c$ remains at the origin and is subject to a rapid change of the mass inside the nucleated bubble. 
This is exactly a kind of quenching in the original cold baryogenesis \cite{Garcia-Bellido:1999xos}, and later, we will discuss the similarity with the winding number production at preheating.   

\section{Dynamics of bubbles
\label{sec:dynamics}}

The nucleation of true vacuum bubbles is followed by their expansion and subsequent collisions.
During the expansion, the bubble walls accelerate to relativistic velocities, and their collisions drive the coupled $\sigma$-Higgs-gauge system into highly nonlinear dynamics. 
In this section, we first characterize the bubble expansion in terms of the wall Lorentz factor and the effective wall width. 
We then present the classical equations of motion governing the coupled system after nucleation and discuss how collisions of bubbles with different Higgs orientations generate the Higgs winding and Chern-Simons number. 
These considerations also identify the characteristic length and time scales relevant for the lattice simulations presented in \SEC{sec:simulation}.

\subsection{Bubble expansion
\label{sec:bubble_expansion}}

One of the characteristic quantities which determine the nature of expanding bubbles is the wall velocity, $v_{\rm w}$.
Depending on their interactions with particles, the expansion mode can be classified into two regimes: runaway and terminal velocity.
The precise estimate of $v_{\rm w}$ is still an ongoing issue and significantly model-dependent (see e.g. Refs.~\cite{Bodeker:2009qy,Bodeker:2017cim,Dorsch:2018pat,BarrosoMancha:2020fay,Laurent:2020gpg,Azatov:2020ufh,Wang:2020zlf,Gouttenoire:2021kjv} for previous studies).
In our analysis, we simply assume a negligibly feeble interaction with the plasma particles with large Lorentz factor $\gamma_*$ at the bubble collision $(v_{\rm w}=1)$.

\begin{figure}[t!]
\centering
\includegraphics[width=7.5cm]{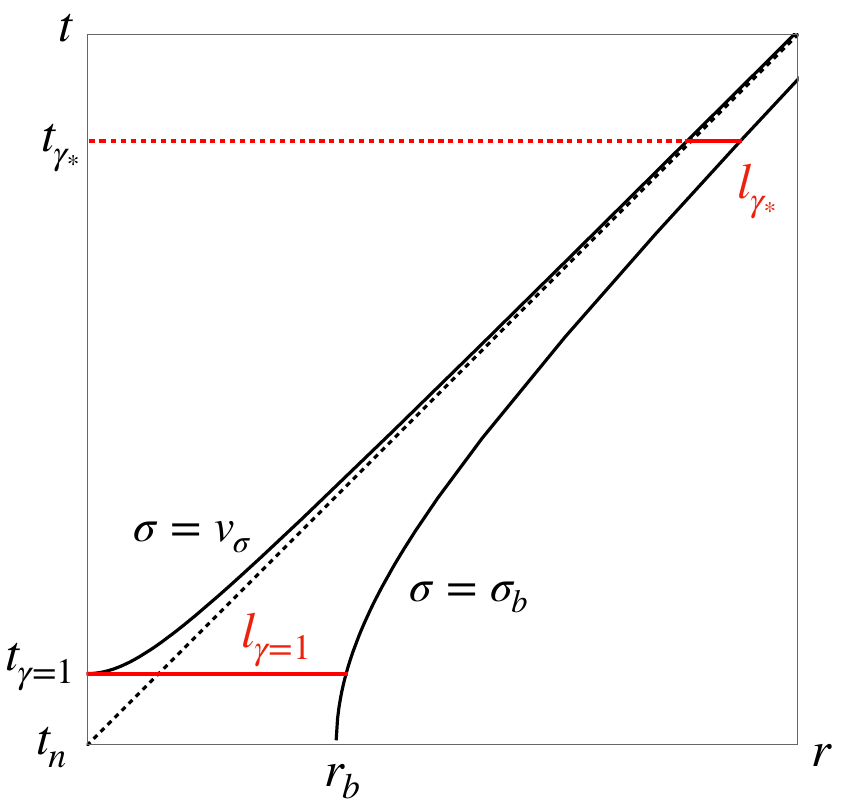}
\caption{
The schematic picture of the evolution of the bubble profile starting at the nucleation $(t=t_n)$.
The black solid lines are the contour lines of $\sigma=\sigma_b,v_\sigma$, and the distance at some time slice is defined as the wall width.
At the collision ($t=t_{\gamma_*}$), the width is given by Eq.~\eqref{eq_distance}.
}
\label{fig:wall}
\end{figure}

Let us study the bubble expansion as a prerequisite for our numerical simulation. 
Since the bubble walls are not accelerated or decelerated by environment, the theory approximately obeys $SO(1,3)$ symmetry before the bubble collisions.
Thus the evolution is described well in terms of $\rho\equiv\sqrt{r^2-t^2}$ and $s\equiv\sqrt{t^2-r^2}$, which are Lorentz-invariant.
\FIG{fig:wall} shows the schematic picture of the field configuration of $\sigma$.
Since we are interested in the case that the Higgs configuration is subdominant $(\epsilon \ll1)$, we focus only on that of $\sigma$.
The contour lines in the space- (time-) like region are given by $\rho~(s)={\rm const}.$. 
In \FIG{fig:wall}, we explicitly show the contour lines for $\sigma=\sigma_b, v_\sigma$.
We define the effective wall width at the collision as the distance between them, given by\footnote{In the thin-wall limit, the wall thickness is set by the inverse mass scale associated with the curvature of the potential around the barrier.
In our case for small $\epsilon$, the thin-wall approximation tends to be broken which does not allow a well-defined wall thickness. Thus, we introduce the effective wall width.}
\beq
\label{eq_distance}
l_{\gamma_*} = \sqrt{r_b^2+t_{\gamma_*}^2}-\sqrt{t_{\gamma_*}^2-t_{\gamma=1}^2} \ .
\eeq
Here we set the nucleation time to $t_n=0$.
As the tunneling point is away from the true minimum in the case of thick wall $(\epsilon\ll 1)$, the scalar fields need time $(=t_{\gamma=1})$ to reach the true minimum.
The duration is estimated by the $SO(1,3)$-symmetric equation \cite{Jinno:2019bxw},
\begin{align}
&\frac{d^2\sigma}{dt^2} +\frac{3}{t}\frac{d\sigma}{dt} +\frac{d V_{\rm eff}}{d\sigma}=0 \ ,
\end{align}
with the initial condition $\sigma=\sigma_0$, where $\sigma_0$ denotes the tunneling point, and $\sigma'=0$ at $m_\sigma t < 1$.
Here $m_\sigma$ denotes the mass at the true vacuum.

Identifying the line of $\sigma=\sigma_b$ as the effective bubble radius $R(t)$, 
we obtain the Lorentz factor at the bubble collision,
\begin{align}
\gamma_* = \frac{R_*}{r_b} \ ,
\label{Lorentz}
\end{align}
with $R_* = R(t_{\gamma_*})$.
By estimating the false vacuum fraction, we obtain the number density of the bubble as \cite{Enqvist:1991xw}
\begin{align}
n_b \simeq \frac{1}{8\pi}\lmk\frac{\beta}{v_{\rm w}}\rmk^3 \ (\simeq R_*^{-3}) \ .
\label{bubble_number}
\end{align}
Here we define the (inverse) duration of phase transition as $\beta\equiv -dS_E/dt|_{t=t_n}$.
Assuming that the Universe is dominated by the vacuum energy, $H\simeq v_\sigma^2/\Mpl$, we obtain
\beq
R_* \simeq (8\pi)^{\frac{1}{3}} v_{\rm w}\frac{\Mpl}{v_\sigma^2}\frac{H}{\beta} \ .
\label{Rstar_beta}
\eeq
The definition of $\gamma_*$ is well-defined ($\gamma_*>1$) only for $R_*\geq r_b$, which indicates the bubbles do not have overlap with each other at the nucleation.
This requires the inequality,
\beq
v_\sigma &\lesssim&(8\pi)^{\frac{1}{3}} \frac{v_{\rm w}}{\beta/H}\frac{\Mpl}{v_\sigma r_b}
\label{well_defined}\\
&\sim& 10^{17}\GeV \cdot \frac{v_{\rm w}}{v_\sigma r_b} \lmk\frac{30}{\beta/H}\rmk. \nonumber
\eeq
For a fixed $\beta/H$, the typical bubble radius at the collision decreases as $R_* \propto v_\sigma^{-2}$ with increasing $v_\sigma$. The critical bubble radius $r_b$ also decreases, but more slowly than $R_*$, since the dimensionless bounce profile is independent of $v_\sigma$. Consequently, the condition $R_* \gtrsim r_b$ imposes an upper bound on the scale of $v_\sigma$. Although this bound is well above the physically motivated values, it becomes relevant when choosing parameters for lattice simulations with finite spatial resolution.

In principle, $\beta$ is determined by the temperature dependence of the Euclidean action $S_E$. Since finite temperature effects are neglected in our model, we do not calculate $\beta$ explicitly and instead assume $\beta/H=\mathcal{O}(10)$ as a representative value for a supercooled phase transition. Verifying whether such a value is realized requires a model-dependent analysis including finite-temperature effects, which is beyond the scope of the present work.

\subsection{Equations of motion}

As the bubbles expand, the subsequent collisions happen and make the system non-linear, which requires a numerical simulation on a lattice spacetime.
Given the initial condition on the bubble profile at the nucleation, the bubble evolution is classically described.
The variational principle leads to the following equations of motion,
\begin{align}
&\ddot{\sigma}-\vec{\nabla}^2\sigma=-\frac{\del V}{\del\sigma} \ , \label{eq:sigma}\\[1ex]
&\ddot{\Phi} -\vec{D}^2\Phi = -\frac{\del V}{\del\Phi^*} \ ,\\[1ex]
&\mathcal{D}_0^{ab} F_{0i}^b -\mathcal{D}_j^{ab}F_{ji}^b = J_i^a \ ,\\[1ex]
&\mathcal{D}_i^{ab}F_{0i}^b =  J_0^a \ , 
\label{eq:Gauss}
\end{align}
where the superscripts $a, b$ are gauge and the subscripts $i,j$ are space indices.
The covariant derivative is given by $\mathcal{D}_\mu^{ab} = \delta^{ab}\del_\mu -g\varepsilon^{abc}A_\mu^{c}$, and the current is defined as
\beq
J_\mu^a = 2g \, \im\left[\Phi^\dagger\frac{\tau^a}{2}D_\mu\Phi\right].
\eeq
The last equation \eqref{eq:Gauss} corresponds to the Gauss law for $SU(2)$.
In the following discussion, we take the temporal gauge, $A_0^a=0$.

The total energy density in the present system is given by
\beq
\rho_{\rm tot} = \left\langle
\sum_f (K_f + G_f) + V
\right\rangle_{\rm vol},
\eeq
where $K_f$ and $G_f$ $(f=\sigma,\Phi,A)$ denote the kinetic and gradient energy for the real scalar, doublet Higgs and $SU(2)$ gauge fields, respectively.
We write them down as follows:
\begin{align}
&K_\sigma =\frac{1}{2}\dot{\sigma}^2 \ ,
~~G_\sigma=\frac{1}{2}\sum_i(\nabla_i\sigma)^2 \ ,\\
&K_\Phi = |\dot{\Phi}|^2 \ ,
~~G_\Phi = \sum_i (D_i\Phi)^\dagger(D_i\Phi) \ ,\\
&K_A = \frac{1}{2}\sum_{a,i} (F^a_{0i})^2 \ ,
~~G_A = \frac{1}{2}\sum_{a,i,j<i}(F_{ij}^a)^2 \ .
\end{align}
We define $\rho_{\rm tot}$ by taking volume average $\langle\cdots\rangle_{\rm vol}$ in a region larger than the spatial scale of particle productions.
When studying on a computer, we can set the volume to the box size larger than the bubble radius at the bubble collision.

The energy density could induce the cosmic expansion. 
However, the phase transition proceeds with time scale faster than the expansion rate, i.e. $\beta/H\gtrsim1$.
Therefore, we ignore the cosmic expansion in our analysis.

\subsection{Asymmetry generation
\label{sec:baryon}}

We consider an $SU(2)$ gauge theory containing $N_F$ flavors of fermions chirally charged under the gauge symmetry and use it as a toy model to study fermion number (or we may call baryon number) generation during a supercooled phase transition. The corresponding fermion number current $j_F^\mu$ is anomalous under $SU(2)$, with its divergence given by
\begin{align}
	    \partial_\mu j_F^\mu = N_{F} \frac{g^2}{32\pi^2} F_{\mu\nu}^a \widetilde F^{a\mu\nu}\, .\label{eq:dark_anomaly}
\end{align}
The anomaly hence directly relates the evolution of the gauge-field topology to a change in the fermion number.

As the bubbles expand in the Universe, numerous collision regions are formed, where Higgs fields with different orientations in the $SU(2)$ space meet.
If the remaining configuration is nontrivial, i.e. $\pi_3(\mathcal{M})=\mathbb{Z}$, we find a gauge texture with topological charges which are called the Higgs winding number $N_w$ and Chern-Simons number $N_{\rm CS}$. They are defined as
\begin{align}
&N_{w} = \frac{1}{24\pi^2}\int d^3x \, \epsilon_{ijk}\Tr[(\del_i\tilde\Phi)\tilde\Phi^\dagger(\del_j\tilde\Phi)\tilde\Phi^\dagger(\del_k\tilde\Phi)\tilde\Phi^\dagger]\label{eq:winding_number} \ ,\\
&N_{\rm CS}(t) -N_{\rm CS}(0) = \frac{g^2}{16\pi^2}\int_0^t dt'\int d^3x \, \Tr[F_{\mu\nu}\widetilde{F}^{\mu\nu}] \ ,\label{eq:CS_number}
\end{align}
where $\tilde\Phi(x)\in SU(2)$ is a $SU(2)$ group element ($2\times 2$ matrix) constructed by the Higgs field $\Phi(x)$, defined as
\begin{align}
    \tilde\Phi(x)\equiv \lmk i\sigma_2\Phi^*,\Phi\rmk/|\Phi| = \begin{pmatrix}
        \Phi_d^* & \Phi_u \\
        -\Phi_u^* & \Phi_d
    \end{pmatrix} /|\Phi|\, . \label{eq:Phi_tilde}
\end{align}
Since the entire space is in the true minimum after the bubble collisions, the remaining energy is stored in the gradients of the Higgs field. The subsequent dynamics is governed by the competition between the unwinding of the Higgs field and the cancellation of the Higgs gradients by the gauge field.
When the texture is larger than $m_W^{-1}$ with $m_W$ the mass of the $SU(2)$ gauge field, the Chern-Simons number changes and finally asymptotes to the Higgs winding number \cite{Turok:1990zg}.
The resulting change in $N_{\rm CS}$ is then converted, through the anomaly in Eq.~\eqref{eq:dark_anomaly}, into a change in the fermion number.

Since our simulation does not include any source of CP violation, positive and negative changes in the Chern-Simons number occur with equal probability, and its expectation value vanishes. Thus, the Chern-Simons number itself cannot be interpreted as a net fermion number asymmetry. Nevertheless, topological transitions lead to fluctuations in the Chern-Simons number that grow with time. We characterize this activity by the variance of the Chern-Simons number and define the corresponding time-dependent diffusion (sphaleron) rate as
\begin{align}
    \Gamma_{\rm CS}(t) \equiv \frac{\dd}{\dd t} \Delta n_{\rm CS}^2(t) \, , \label{eq:CS_diffusion_rate}
\end{align}
with the ensemble variance defined as
\begin{align}
\Delta n_{\rm CS}^2(t)\equiv \left\langle n_{\rm CS}^2(t)\right\rangle_{\rm ens}-\left\langle n_{\rm CS}(t)\right\rangle_{\rm ens}^2 \, .
\end{align}
Here $n_{\rm CS}\equiv N_{\rm CS}/V$ and $\langle\cdots\rangle_{\rm ens}$ denotes the average over different realizations of random bubble positions and Higgs orientations.
A nonzero asymmetry is then estimated by introducing CP violation perturbatively and treating its effect as a small bias.

We introduce a dimension-6  CP-violating operator,
\begin{align}
    \mathcal L_{\rm CPV} = \delta_{\rm CP}\frac{\Phi^\dagger\Phi}{\Lambda_{\rm CP}^2}  \frac{g^2}{16\pi^2} \Tr\lkk F_{\mu\nu}\widetilde F^{\mu\nu}\rkk \, ,\label{eq:cpv_operator}
\end{align}
where $\delta_{\rm CP}$ is a dimensionless coefficient and
$\Lambda_{\rm CP}$ denotes the cutoff scale at which the operator is
generated.
According to Eq.~\eqref{eq:dark_anomaly}, this CP-violating operator can be rewritten as
\begin{align}
    \mathcal L_{\rm CPV} = \delta_{\rm CP}\frac{\Phi^\dagger\Phi}{N_F\Lambda_{\rm CP}^2} \del_\mu j_{F}^\mu = - \frac{\delta_{\rm CP}}{N_F\Lambda_{\rm CP}^2} \del_\mu\lmk\Phi^\dag\Phi \rmk j^\mu_{F} \, ,\label{eq:cpv_operator_integrated}
\end{align}
where we have performed the integration by parts in the last equality.
After the supercooled phase transition, the Universe can be considered to be approximately homogeneous.
Then, for a spatially averaged, time-dependent scalar condensate $\overline{\Phi^\dag\Phi}(t)\equiv \frac{1}{L^3}\int_V \dd^3 x \, \Phi^\dag\Phi(x)$, the temporal part of Eq.~\eqref{eq:cpv_operator_integrated} acts as a chemical potential,
\begin{align}
    \mu_F(t) \equiv \frac{\delta_{\rm CP}}{N_F\Lambda_{\rm CP}^2}\frac{\dd}{\dd t}\overline{\Phi^\dag\Phi}\, . \label{eq:chemical_potential}
\end{align}
The volume-averaged fermion number density can be described by a Boltzmann-like equation,
\begin{align}
	    \frac{\dd n_F}{\dd t} = \Gamma_{\rm CS}(t) \lkk \frac{\mu_F(t)}{T_{\rm eff}(t)} - \mathcal A n_F(t) \rkk \, , \label{eq:dark_number_boltzmann}
\end{align}
where the time-dependent Chern-Simons diffusion rate is defined in Eq.~\eqref{eq:CS_diffusion_rate}, $\mathcal A$ parametrizes the washout effect
and $T_{\rm eff}$ denotes the effective temperature for the diffusion rate, which is estimated as $T_{\rm eff}(t) = \frac{\Gamma_{\rm CS}^{1/4}(t)}{\alpha_W}$ with $\alpha_W=g^2/(4\pi)$
\cite{Bhusal:2025lvm}.
In the following estimate we neglect the washout effect. Then, the generated fermion number density is obtained as
\begin{align}
    n_F &= \int \dd t\,\Gamma_{\rm CS}(t) \frac{\mu_F(t)}{T_{\rm eff}(t)} = \alpha_W \int \dd t\, \Gamma_{\rm CS}^{3/4}(t)\mu_F(t) \notag \\[1ex]
    &=\frac{\alpha_W\delta_{\rm CP}}{N_F\Lambda_{\rm CP}^2} \int\dd t\,\Gamma_{\rm CS}^{3/4}(t) \frac{\dd}{\dd t} \overline{\Phi^\dagger\Phi}(t) \ .
    \label{eq:dark_density_effective}
\end{align}
The entropy density at reheating after the phase transition is given by
\begin{align}
    s(T_{\rm rh}) = \frac{2\pi^2}{45} g_{*}(T_{\rm rh})T_{\rm rh}^3 \, , \label{eq:entropy_density}
\end{align}
where $g_*$ denotes the effective number of relativistic degrees of freedom, and $T_{\rm rh}$ is the reheating temperature
after the supercooled phase transition.
The fermion number yield can be finally estimated as
\begin{align}
    Y_F \equiv \frac{n_F}{s} = \frac{45}{2\pi^2g_{*}T_{\rm rh}^3} \frac{\delta_{\rm CP}\alpha_W}{N_F\Lambda_{\rm CP}^2} \int \dd t\, \Gamma_{\rm CS}^{3/4}(t) \frac{\dd}{\dd t} \overline{\Phi^\dag\Phi}(t)\, .
    \label{yield}
\end{align}

In the present work, we estimate the generated asymmetry in a supercooled phase transition by lattice simulations.
Ref.~\cite{Bhusal:2025lvm} showed that the Chern-Simons number can be efficiently generated for a nearly degenerate potential ($\epsilon \sim 1$). However, for $\epsilon \ll 1$, efficient production becomes difficult because the Higgs field does not bounce back and therefore does not pass through zero.
In our setup, the Higgs sector $(\Phi)$ is separated from the phase transition sector $(\sigma)$, and thus, the Higgs field can be zero at the collision even for $\epsilon\ll1$.\footnote{
Here, we implicitly assume that the Higgs field has reached the true-vacuum minimum by the time the bubbles collide. The collision can then induce sufficiently large Higgs-field oscillations for the field to reach zero. This condition can be readily satisfied, as long as $\lambda_\Phi\gtrsim 10^{-32}(v_\sigma/1\GeV)^2(\beta/H/100)^2$.
}

\begin{table}[!t]
    \centering
    \begin{tabular}{c|ccc}
        \hline
         & $\epsilon$ & $a$ & $b$ \\
        \hline
        BP1 & $10^{-2}$ & $0.3$ & $0.1631$  \\
        BP2 & $10^{-3}$ & $0.3$ & $0.0223$  \\
        BP3 & $10^{-4}$ & $0.3$ & $-0.1639$ \\
        \hline
    \end{tabular}
    \caption{
        The benchmark for dimensionless parameters in the scalar potential, where $\epsilon$ is determined by $a,b$ in Eq.~\eqref{degeneracy}.
        As mentioned in Sec.~\ref{sec:Setup}, we fix $\lambda_\sigma=\lambda_\Phi=1$ and $v_\Phi=0.8 v_\sigma$.
    }
    \label{tab:dimless_pot_benchmarks}
\end{table}

\section{Simulation on lattice
\label{sec:simulation}}

We now turn to the numerical simulations of the bubble dynamics and the associated Higgs-gauge evolution. Using the bounce configurations obtained in section~\ref{vacuum transition} as initial conditions, we solve Eqs.~\eqref{eq:sigma}-\eqref{eq:Gauss} on a three-dimensional lattice and follow the real-time evolution via bubble expansion and collisions. We employ the gauge-invariant lattice formulation for scalar and non-Abelian gauge fields developed in Ref.~\cite{Figueroa:2020rrl}, which allows us to study the generation of the Higgs winding and Chern-Simons number in the highly nonlinear regime. The lattice gauge formulation employed in our simulations is summarized in Appendix \ref{app:lattice}.
We first discuss the simulation parameters and the requirements imposed by the finite spacetime resolution, before specifying the initial conditions and presenting the numerical results.

\subsection{Parameters for simulation
\label{sec:lattice_setup}}

\begin{table*}[!t]
    \centering
    \begin{tabular}{c @{\hspace{12pt}}c@{\hspace{12pt}}ccc
    @{\hspace{12pt}}c@{\hspace{12pt}}ccc
    @{\hspace{12pt}}c@{\hspace{12pt}}ccc}
        \hline\hline
    	\noalign{\vskip 3pt}
    	& \multicolumn{4}{c}{BP1: $\epsilon=10^{-2}$}
    	& \multicolumn{4}{c}{BP2: $\epsilon=10^{-3}$}
    	& \multicolumn{4}{c}{BP3: $\epsilon=10^{-4}$} \\[3pt]
    \cline{2-13}
    \noalign{\vskip 4pt}
    $\gamma_*$
    & $Lm_\sigma$ & $l_{\gamma_*}m_\sigma$ & $N$ & $v_\sigma\,[10^{15}\,\mathrm{GeV}]$
    & $Lm_\sigma$ & $l_{\gamma_*}m_\sigma$ & $N$ & $v_\sigma\,[10^{15}\,\mathrm{GeV}]$
    & $Lm_\sigma$ & $l_{\gamma_*}m_\sigma$ & $N$ & $v_\sigma\,[10^{15}\,\mathrm{GeV}]$ \\[3pt]
    \hline
        $4.0$ & $181.89$ & $1.63$ & $224$
              & $8.07$ & $183.26$ & $1.66$
              & $222$ & $9.66$ & $219.11$ & $1.94$
              & $226$ & $9.36$ \\
        $4.5$ & $204.63$ & $1.44$ & $284$
              & $7.18$ & $206.16$ & $1.47$
              & $282$ & $8.56$ & $246.50$ & $1.71$
              & $288$ & $8.32$ \\
        $5.0$ & $227.36$ & $1.30$ & $352$
              & $6.46$ & $229.07$ & $1.32$
              & $348$ & $7.73$ & $273.88$ & $1.54$
              & $356$ & $7.49$ \\
        $5.5$ & $250.10$ & $1.18$ & $426$
              & $5.87$ & $251.98$ & $1.20$
              & $422$ & $7.02$ & $301.27$ & $1.40$
              & $432$ & $6.81$ \\
        $6.0$ & $272.83$ & $1.08$ & $508$
              & $5.38 $ & $274.88$ & $1.09$
              & $504$ & $6.44$ & $328.66$ & $1.28$ & $514$ & $6.24$\\
        \hline\hline
    \end{tabular}
    \caption{
        Simulation settings for the three benchmarks.
        Each simulation uses a cubic lattice with $N^3$ sites. 
        The lattice size $Lm_\sigma$, the bubble-wall width $l_{\gamma_*}$ and $v_\sigma$ are specified separately for each $\gamma_*$.
        The required number of grids is then determined by $N=L/(c_x l_{\gamma_*})$.
        We fix the number of bubbles $N_b=10$ and the dimensionless inverse transition duration $\beta/H=30$.
    }
    \label{tab:simulation_settings}
\end{table*}

First, let us find parameter sets suitable for the simulation with sufficient resolution of spacetime.
After the nucleation of the bubbles, the fields start to evolve at the time scale of $m_\sigma^{-1}$.
To resolve the time evolution, we need to take the step size $\delta t$ small enough:
\beq
m_\sigma\delta t < c_t \ , 
\label{time_resolution}
\eeq
where $c_t < 1$ is a numerical parameter; see Appendix~\ref{app:sys_uncertanties} for details.
Since it takes $R_*$ for the bubbles to collide, the number of steps in simulations can be estimated as 
\beq
N_s \sim \frac{R_*}{\delta t} \ .
\eeq
\EQ{time_resolution} is then rephrased as a lower bound on $v_\sigma$,
\beq
v_\sigma &\gtrsim&\frac{(8\pi)^{1/3}}{N_s c_t} v_{\rm w} \frac{m_\sigma}{v_\sigma}\frac{H}{\beta}\Mpl\\
&\sim& 10^{15}\GeV \, v_{\rm w} \frac{m_\sigma}{v_\sigma}\lmk\frac{10^{3}}{N_s}\rmk\lmk\frac{0.1}{c_t}\rmk\lmk\frac{30}{\beta/H}\rmk.
\nonumber \eeq
If the maximum step number is taken to be $\mathcal{O}(1000)$, we need to take such a large value of $v_\sigma$ for a small $\beta/H$.
Together with the upper bound (\ref{well_defined}), we only have a limited range of $v_\sigma$ for $\beta/H\sim \mathcal{O}(10)$,
\beq
10^{15}\GeV \lesssim v_\sigma \lesssim 10^{17}\GeV \ .
\eeq
In addition to the time resolution, one needs to consider the resolution of space.
The Higgs winding number can be produced only if more than two bubbles collide due to its topological nature, which has been confirmed in Ref.~\cite{Bhusal:2025lvm}. 
In order to simulate the production of the winding number, the simulation volume must contain multiple bubbles at the time of collision. We therefore choose the box size $L$ according to the geometric relation,
\beq 
\lmk \frac{\sqrt{3}}{2}L\rmk^3 \simeq N_b\times \frac{4\pi}{3}R_*^3 \ ,
\eeq
where $R_*$ is the typical bubble radius at collision and $N_b$ is the number of bubbles in the simulation volume.
Since $R_*$ is set by $\beta^{-1}$ through Eq.~\eqref{Rstar_beta}, the typical
separation between bubbles and their radius at collision are controlled
by $\beta$. In contrast, the characteristic microscopic scales of an
individual bubble, such as its initial radius and wall width, are set
by $v_\sigma$. The system therefore involves two distinct scales:
the macroscopic scale associated with the bubble distribution, determined
by $\beta$, and the microscopic scale associated with the bubble profile,
determined by $v_\sigma$.
Among these scales, the bubble-wall width is the smallest spatial scale
that must be resolved on the lattice. We have confirmed that the estimated
Chern-Simons number converges when the lattice spacing is sufficiently
small compared with the effective wall width at collision. This requires
(see Appendix \ref{app:sys_uncertanties})
\beq
\delta x \equiv \frac{L}{N} \lesssim l_{\gamma_*}/c_x \ ,
\label{space_resolution}
\eeq
with $c_x=2$.
Note that for fixed $\beta/H$, decreasing $v_\sigma$ increases the macroscopic bubble scale relative to the microscopic wall width. Consequently, resolving the bubble wall becomes increasingly demanding for smaller $v_\sigma$. The spatial-resolution condition in Eq.~\eqref{space_resolution} therefore imposes a lower bound on $v_\sigma$, which is comparable to or stronger than that obtained from the time-resolution condition in Eq.~\eqref{time_resolution}.

As discussed in \SEC{sec:bubble_nucleation}, completion of bubble nucleation in the
phenomenologically motivated range
$v_\sigma \sim 100~{\rm MeV} - 100~{\rm TeV}$ typically requires
$a=\mathcal{O}(0.1)$ and $\epsilon\sim10^{-4} - 10^{-2}$.
For a slow transition with $\beta/H=\mathcal{O}(10)$,
however, the large hierarchy between the macroscopic bubble scale
and the microscopic wall scale makes a direct simulation of this
parameter region computationally inaccessible.
We therefore adopt the bounce profiles obtained for the
phenomenologically motivated values of $a$ and $\epsilon$, while
rescaling the overall scale $v_\sigma$ to larger values that can be
resolved on the lattice.
The resulting parameter sets should not be interpreted as
self-consistent models of bubble nucleation at these large values of
$v_\sigma$; rather, they provide numerically tractable initial
conditions that retain the dimensionless bubble profiles relevant
to the physical parameter region.
The subsequent bubble evolution depends also on the hierarchy
between the initial bubble size and the typical bubble separation,
which is characterized by the Lorentz factor $\gamma_*$ at collision.
We therefore study the dependence of the bubble dynamics and the
resulting winding and Chern-Simons number production on
$\epsilon$ and $\gamma_*$.\footnote{Alternatively, we could take a very large $\beta$ with a natural value of $v_\sigma$ \cite{Bhusal:2025lvm}.
However, the geometric relation of bubbles is the same.}

We consider three benchmark scalar potentials, whose dimensionless parameters are listed in Table~\ref{tab:dimless_pot_benchmarks}.
For each benchmark, we fix the gauge coupling $g$, the number of bubbles $N_b$ and the dimensionless inverse transition duration $\beta/H$.
The VEV of the phase transition field $v_\sigma$ is then determined by the chosen Lorentz factor at the collision time $\gamma_*$ through Eq.~\eqref{Rstar_beta}.
The resulting simulation parameters are summarized in
Table~\ref{tab:simulation_settings}.
These parameters satisfy the constraints from the size of nucleated bubble (\ref{well_defined}) and from spacetime resolution (\ref{time_resolution}), (\ref{space_resolution}).
With a fixed $\beta/H=30$, increasing $\gamma_*$ corresponds to
decreasing $v_\sigma$, allowing us to investigate the dependence
on the phase transition scale and explore extrapolations beyond
the range accessible to our simulations.
In addition, we fix
\begin{equation}
\begin{aligned}
    &\delta_{\rm CP}=1\, , \quad g_*=106.75\, , \quad N_F=3\, , \\[1ex]
    &\Lambda_{\rm CP}\simeq v_\sigma\, , \quad T_{\rm rh}\simeq 0.3v_\sigma\, .
\end{aligned}
\end{equation}
With these choices, we estimate the fermion asymmetry from the Chern-Simons dynamics.

\begin{figure}[!t]
    \centering
    \includegraphics[width=0.9\linewidth]{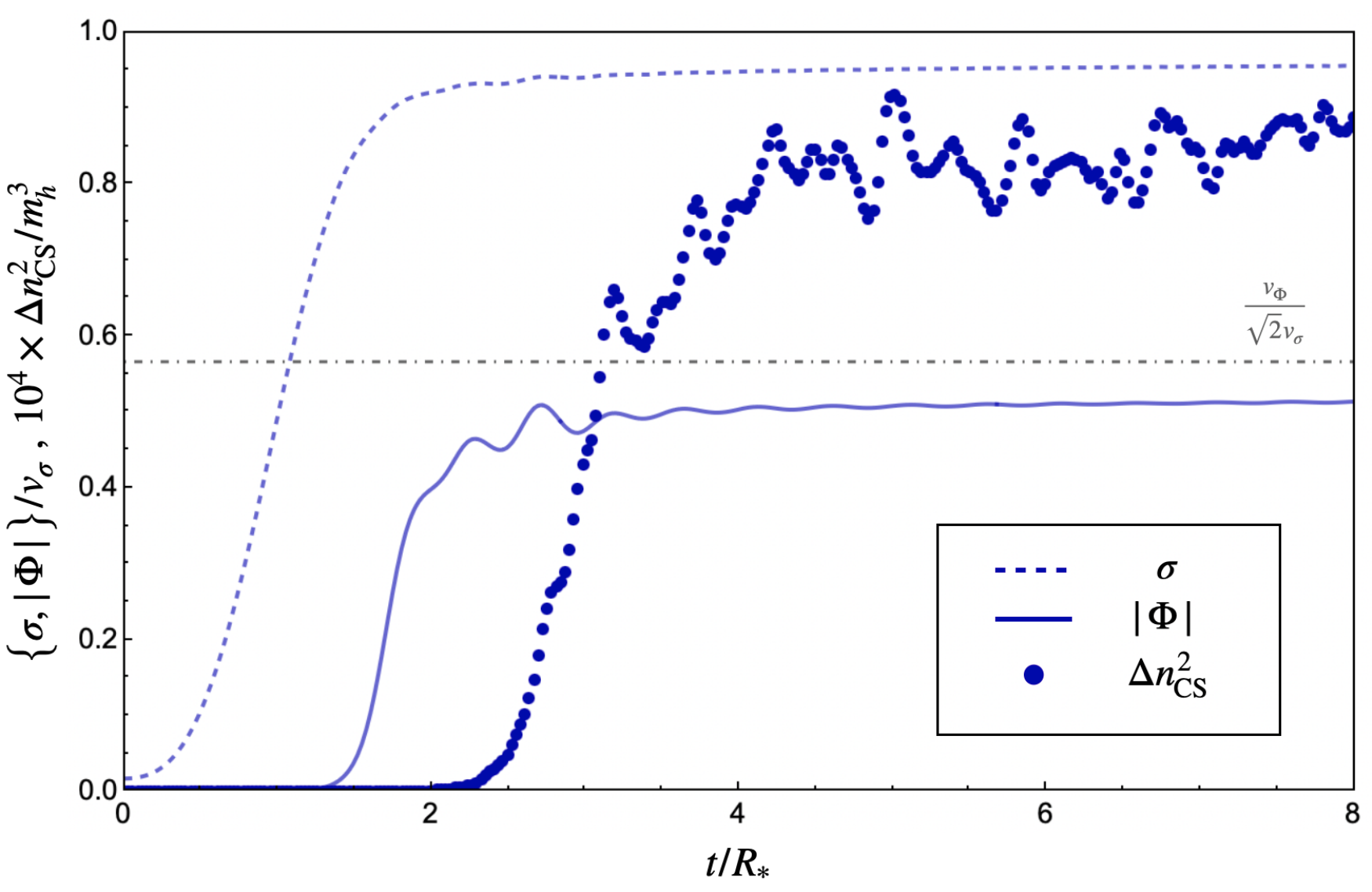}
    \caption{The time evolution of the field values and the Chern-Simons number variance obtained from an ensemble of simulations with $\epsilon=10^{-3},\gamma_*=4.5$.
    The solid and dashed lines represent the spatially and ensemble-averaged values of $|\Phi|$ and $\sigma$ respectively, and both are normalized by $v_\sigma$.
    The dot dashed line denote the VEV of the Higgs field.
    The points show the Chern-Simons number variance, normalized by the Higgs mass and rescaled by a factor of $10^4$.}
    \label{fig:time_evolution}
\end{figure}

\subsection{Initial conditions}

We construct the initial field configuration by placing a fixed number of nucleated bubbles within the simulation volume.
The center positions of bubbles are chosen randomly, while the initial bubble profile is taken from the bounce solution obtained in Sec.~\ref{sec:bubble_nucleation}. 
More explicitly, for a bubble nucleated at ${\bf x}_n$, the scalar profiles are assigned as
\begin{align}
\sigma({\bf x})=\sigma_c(|{\bf x}-{\bf x}_n|)\, ,\qquad |\Phi({\bf x})|=\frac{h_c(|{\bf x}-{\bf x}_n|)}{\sqrt{2}}\, ,
\end{align}
where $\sigma_c$ and $h_c$ denote the $O(4)$-symmetric bounce profiles at $\tau=0$. 
For multiple bubbles, we superpose the individual profiles in such a way that the field value approaches the true-vacuum value inside each bubble and the false-vacuum value outside all bubbles. 
The orientation of the Higgs field in the $SU(2)$ space is chosen independently for each bubble. 
Namely, inside the $n$-th bubble, we take
\begin{align}
\Phi_n({\bf x})=U_n\begin{pmatrix}0\\ h_c(|{\bf x}-{\bf x}_n|)/\sqrt{2}\end{pmatrix}\, ,
\end{align}
where $U_n\in SU(2)$ is a randomly generated group element. 

The initial time derivatives of the scalar fields are determined by the classical continuation of the bounce solution. Since the bubbles are nucleated from the $O(4)$-symmetric configuration at $\tau=0$, the scalar velocities vanish at the nucleation time,
\begin{align}
\dot{\sigma}({\bf x})=0 \ ,\qquad \dot{\Phi}({\bf x})=0 \ .
\end{align}
The gauge fields and their canonical momenta are also set to zero initially,
\begin{align}
A_i^a({\bf x})=0 \ ,\qquad \dot A_i^a({\bf x})=0 \ .
\end{align}
With this choice, the Gauss constraint \eqref{eq:Gauss} is automatically satisfied at the initial time.

\begin{figure}[t!]
    \centering
    \includegraphics[width=0.9\linewidth]{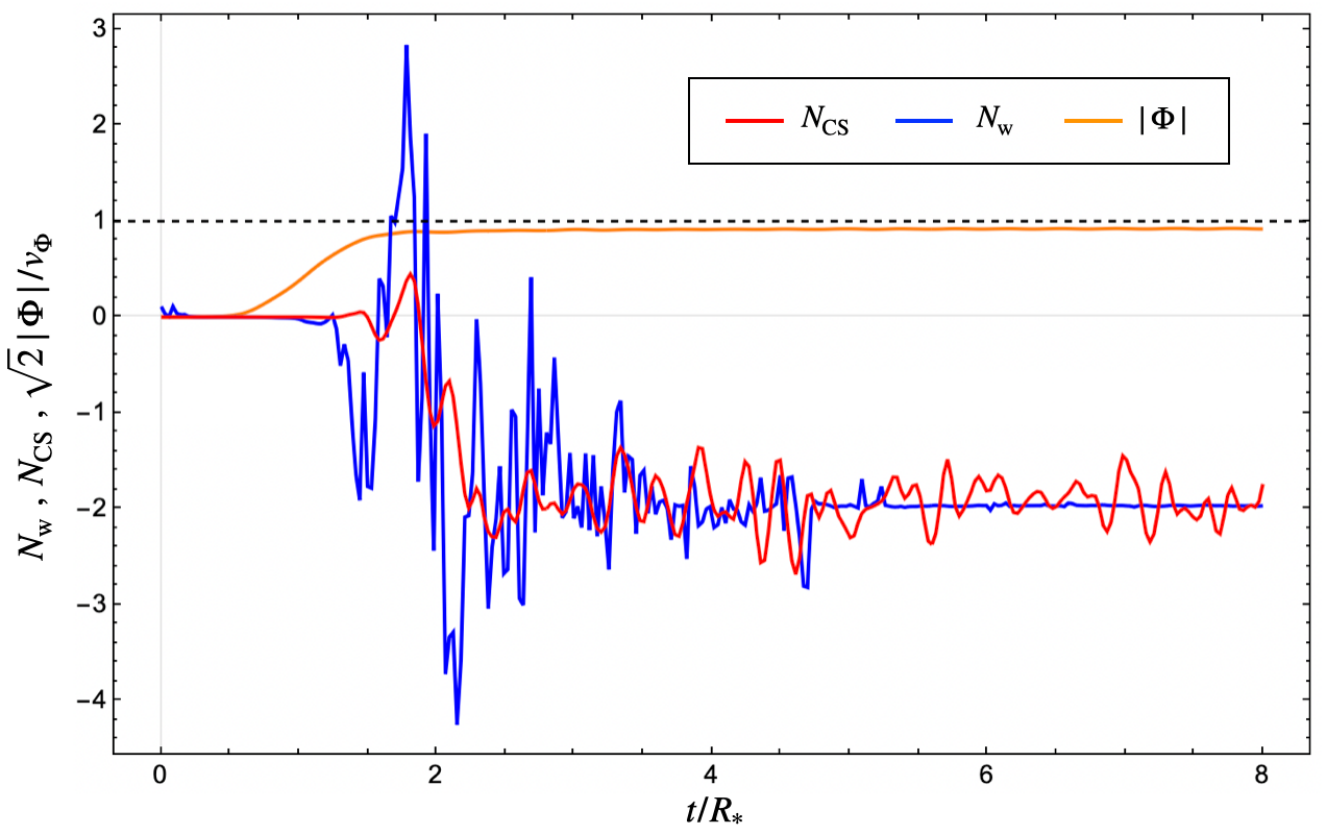}
    \caption{The time evolution of the Chern-Simons number and the winding number in one simulation of the $\epsilon=10^{-2},~\gamma_*=4.5$ case.
    The red and blue solid lines represent the Chern-Simons number and the Higgs winding number, respectively.
    The orange solid line denotes the absolute value of the Higgs field and the black dashed line shows its field value at the true vacuum.}
    \label{fig:winding_CS}
\end{figure}

\subsection{Results}
As shown in Table ~\ref{tab:dimless_pot_benchmarks}, we fix the scale-invariance breaking parameter $a=0.3$ and take three representative values of the degeneracy parameter, $\epsilon=10^{-2},10^{-3},10^{-4}$. 
The gauge coupling is taken, in most cases, to that of the SM $SU(2)_L$, $g=0.652$.
It will be varied when we study the power spectrum of the Chern-Simons variance.
We also vary the Lorentz factor at collision, $\gamma_*$, by changing the physical lattice size while keeping the other parameters fixed.
For each benchmark point, we perform an ensemble of simulations with different random bubble positions and Higgs orientations. 
This allows us to extract the statistical behavior of the generated fermion number and its dependence on the phase transition parameters.

\subsubsection{Evolution}
We first investigate the real-time evolution of the scalar fields. 
The solid and dashed lines in Fig.~\ref{fig:time_evolution} show the volume-averaged field values of $\sigma$ and $|\Phi|$ as functions of time normalized by the typical bubble radius at collision, $R_*$. 
The singlet field $\sigma$ starts from the false vacuum and approaches the true vacuum as the bubbles expand and collide. 
After the collision, $\sigma$ overshoots the true minimum and begins to oscillate around it.\footnote{The plotted dashed curve in Fig.~\ref{fig:time_evolution} represents the spatially averaged field value of $\sigma$ rather than its local evolution. 
The smooth behavior of the averaged field can therefore result from the suppression of local oscillations and overshoots by averaging over regions with different collision times and oscillation phases.}
These oscillations are important because they induce a time-dependent effective mass for the Higgs field through the coupling in Eq.~\eqref{potential}.
The Higgs field remains small before the bubble collision, especially for small $\epsilon$, because the exit point of the bounce solution in the Higgs direction is suppressed. 
After the collision, however, the oscillation of $\sigma$ can drive the Higgs field away from the origin. As a result, the Higgs field grows rapidly and approaches its true-vacuum value.

Fig.~\ref{fig:time_evolution} also shows the generated Chern-Simons variance, indicated by the scattered points.
Efficient Chern-Simons number production begins only when the Higgs field approaches its true-vacuum value, after which the Chern-Simons variance quickly saturates.
As mentioned in Sec.~\ref{sec:baryon}, the generation of the Chern-Simons number is always accompanied by the generation of the Higgs winding number.
In Fig.~\ref{fig:winding_CS} we show the time evolution of the Chern-Simons number and Higgs winding number in one representative simulation with $\epsilon=10^{-2},~\gamma_*=4.5$.
The winding number exhibits large oscillations initially as the Higgs field grows towards its VEV, reflecting the strongly non-linear dynamics.
The Chern-Simons number broadly follows the trajectory of the winding number but has smaller oscillation amplitude.
At late times, $t>6R_*$, both of them settle into the same topological sector, $N_w\simeq N_{\rm CS}\simeq -2$.
The Chern-Simons number has more noise because Eq.~\eqref{eq:CS_number} is a total derivative and a naive lattice discretization fails to retain this topological property.

\subsubsection{Spectrum}

\begin{figure}[!t]
    \centering
    \includegraphics[width=0.9\linewidth]{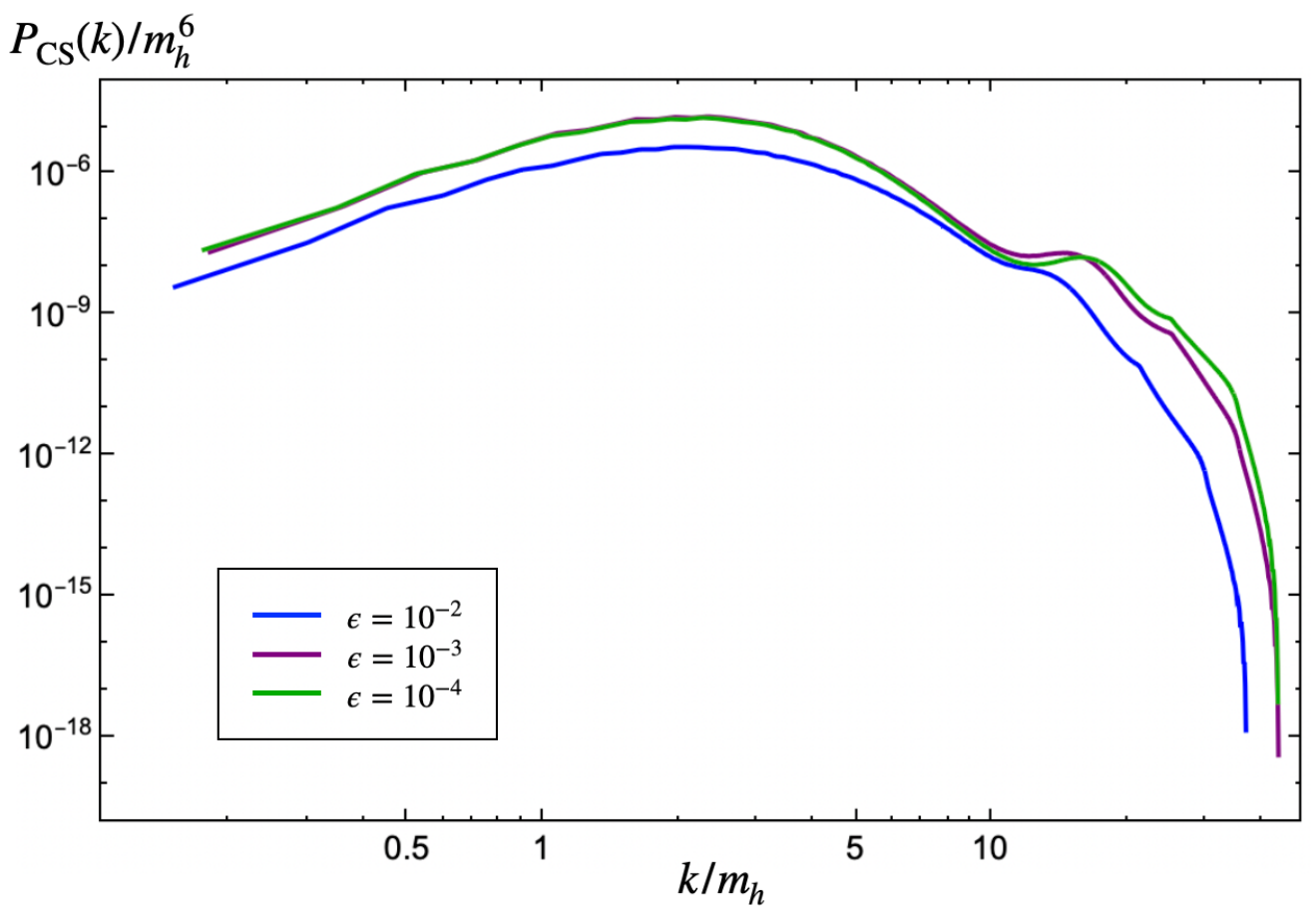}
    \caption{Power spectrum of the Chern-Simons number density for $\epsilon=\{10^{-2},10^{-3},10^{-4}\},\gamma_*=4.5$.
    The blue, purple and green solid lines represent the cases of $\epsilon=10^{-2},10^{-3},10^{-4}$, respectively.}
    \label{fig:CS_spectrum_epsilon}
\end{figure}

Let us next calculate the power spectrum of the Chern-Simons number density. 
Defining the Fourier transform of the Chern-Simons number density as,
\begin{align}
    n_{\rm CS}({\bf k}) = \int\dd^3 x e^{-i\mathbf{k}\cdot \mathbf{x}}n_{\rm CS}({\bf x})\, ,
\end{align}
the spectrum is obtained through the ensemble-averaged two-point function,
\begin{align}
\left\langle n_{\rm CS}({\bf k})n_{\rm CS}({\bf k}’)\right\rangle_{\rm ens}=(2\pi)^3\delta^{(3)}({\bf k}+{\bf k}’) \frac{2\pi^2}{k^3}P_{\rm CS}(k) \ .
\end{align}
The numerical result for $\epsilon=\{10^{-2},10^{-3},10^{-4}\},\gamma_*=4.5$ is shown in Fig.~\ref{fig:CS_spectrum_epsilon}.
The spectra for $\epsilon=10^{-3}$ and $10^{-4}$ have larger amplitudes than that for $\epsilon=10^{-2}$, indicating enhanced fluctuations in the Chern-Simons number density  and a larger Chern-Simons variance at smaller $\epsilon$. 
Despite the different values of $m_\sigma$ and $R_*$, the three spectra peak at approximately the same momentum, $k_{\rm peak}\sim m_h$. 
This suggests that the characteristic correlation length of the Chern-Simons number density is set by the Higgs-sector scale, $\ell_{\rm CS}\sim k_{\rm peak}^{-1}\sim m_h^{-1}$.

\begin{figure}[t!]
    \centering
    \includegraphics[width=0.9\linewidth]{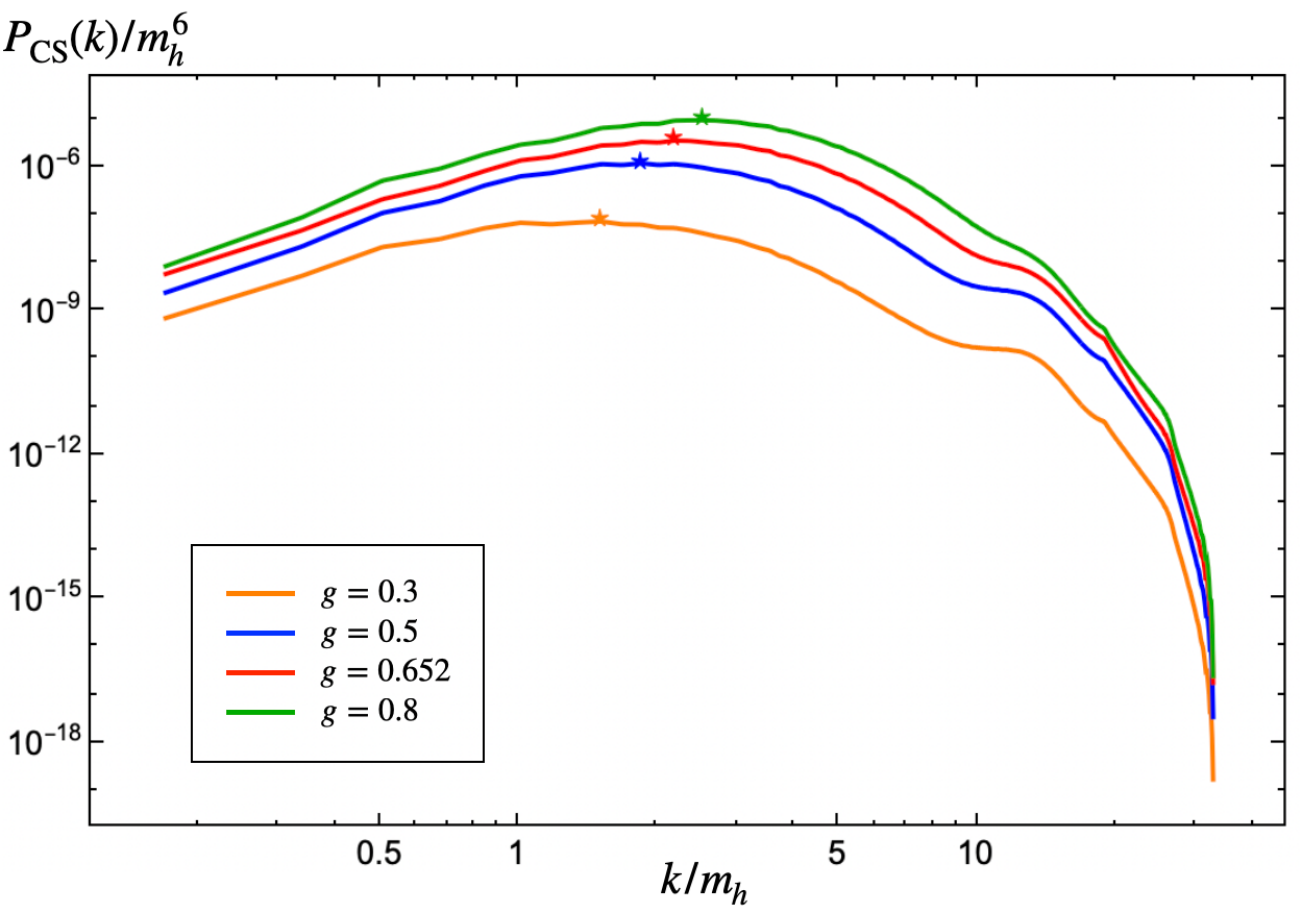}
    \caption{Power spectrum of the Chern-Simons number density for $g=\{0.3,0.5,0.652,0.8\},\epsilon=10^{-2}$ and $\gamma_*=4$.
    The orange, blue, red and green solid lines represent the cases of $g=0.3,0.5,0.652,0.8$, respectively. The stars put on the solid curves show the peaks of the power spectrum.}
    \label{fig:CS_spectrum_g}
\end{figure}

To be more precise, one expects the relevant length scale to be related to the inverse gauge boson mass $m_W^{-1}$. 
As discussed in Sec.~\ref{sec:baryon}, whether a Higgs texture develops into a gauge texture or simply decays depends on its size relative to the gauge-field screening length $m_W^{-1}$.
To test this interpretation, we vary the gauge coupling $g$ while fixing $\epsilon=10^{-2}$ and $\gamma_*=4$, and perform ensembles of simulations with different initial configurations for each value of $g$.
The resulting power spectra are shown in Fig.~\ref{fig:CS_spectrum_g}.
As $g$, and consequently $m_W$, increases, the peak shifts toward larger momenta.
This behavior is consistent with the physical expectation that the characteristic Chern-Simons correlation length decreases as $m_W^{-1}$.
The increase in spectral amplitude with $g$ is expected: as $m_W^{-1}$ decreases, a larger fraction of Higgs textures tends to be sufficiently large to induce gauge-field winding, thereby enhancing the Chern-Simons number variance.

\begin{figure*}[t!]
\centering

\begin{minipage}[c]{0.65\textwidth}
\centering

\subfloat[]{%
    \includegraphics[width=0.485\linewidth]{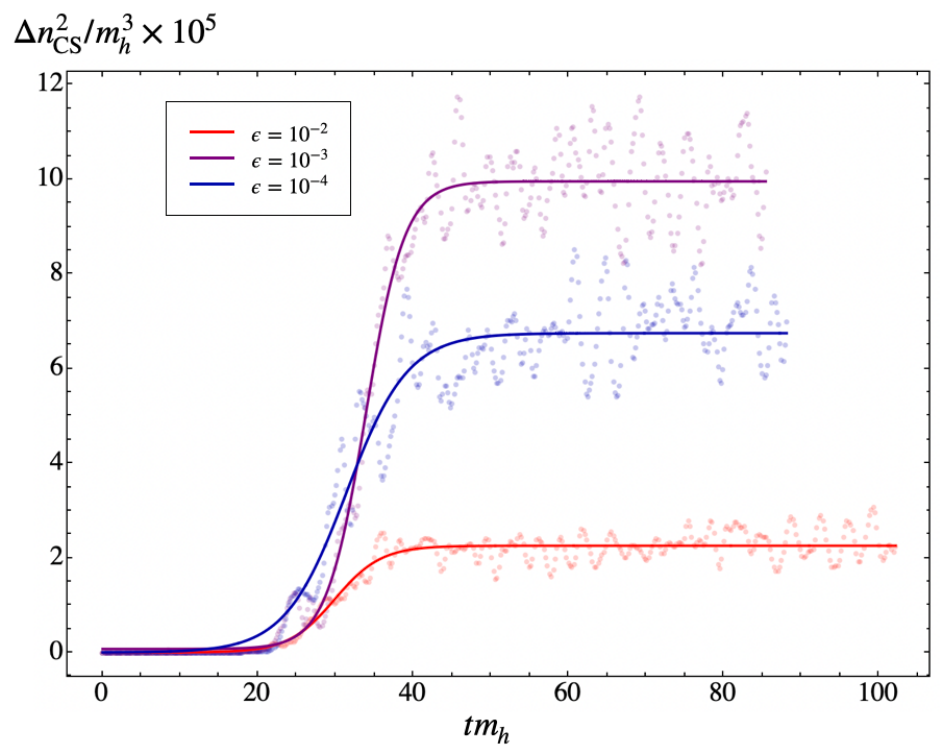}%
    \label{fig:CS-variance-fit}%
}\hfill%
\subfloat[]{%
    \includegraphics[width=0.485\linewidth]{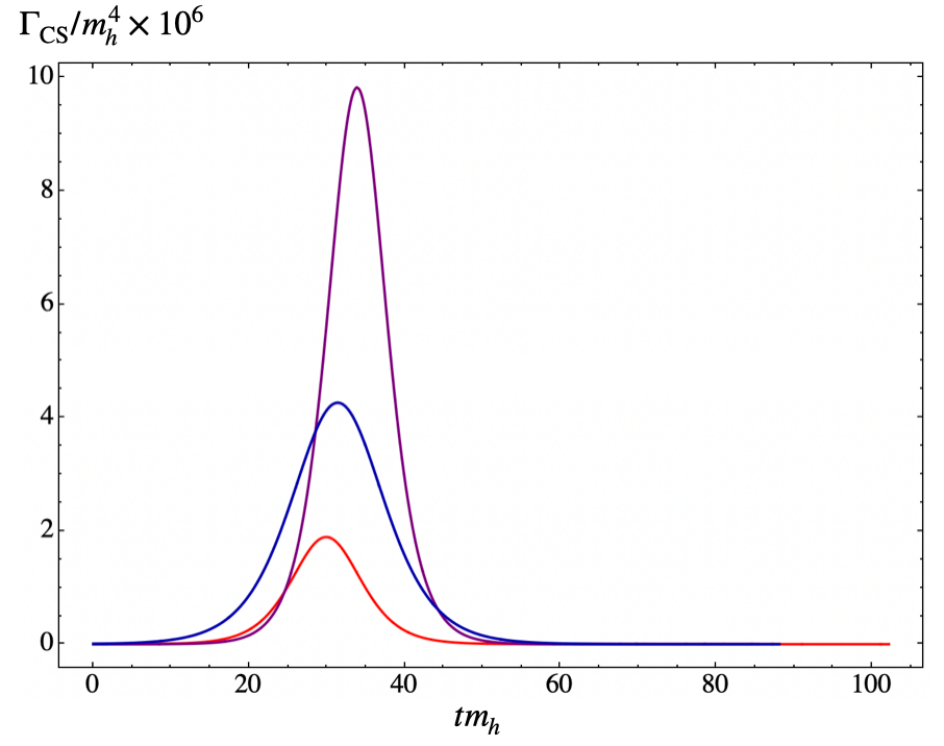}%
    \label{fig:Gamma-CS}%
}

\par\vspace{1mm}

\subfloat[]{%
    \includegraphics[width=0.485\linewidth]{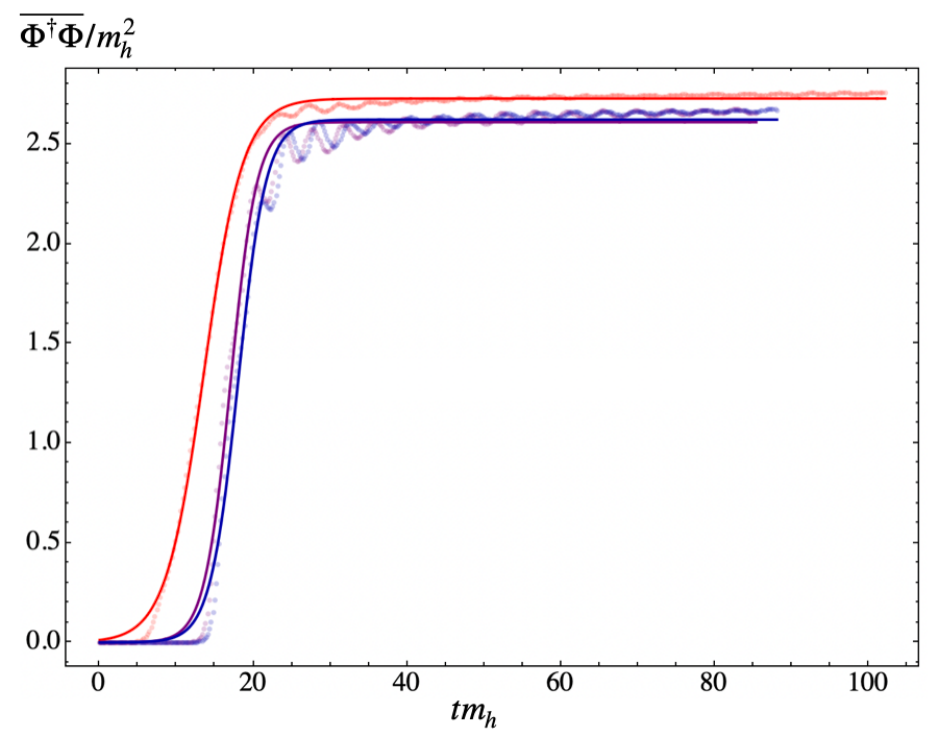}%
    \label{fig:Phi-fit}%
}\hfill%
\subfloat[]{%
    \includegraphics[width=0.485\linewidth]{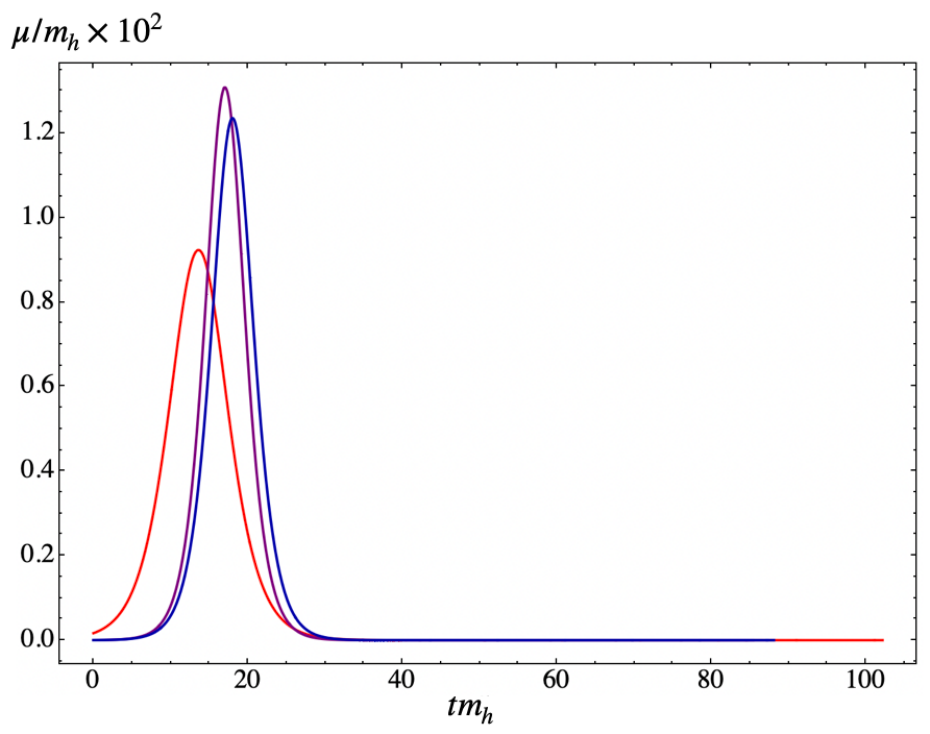}%
    \label{fig:chemical-potential}%
}

\end{minipage}%
\hfill%
\begin{minipage}[c]{0.33\textwidth}
\centering

\subfloat[]{%
    \includegraphics[width=0.97\linewidth]{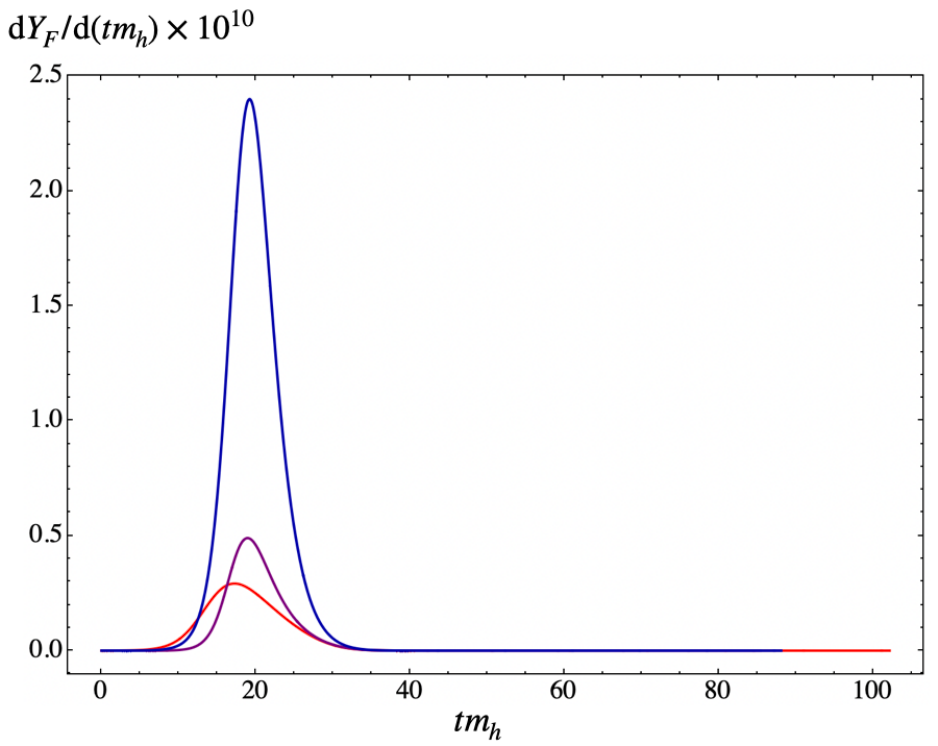}%
    \label{fig:baryon-yield-integrand}%
}

\end{minipage}

\caption{%
Time evolution of the quantities entering the estimate of the generated
asymmetry. The red, purple, and blue solid curves correspond to
$\epsilon=10^{-2}$, $10^{-3}$, and $10^{-4}$, respectively.
Panel (a) shows the Chern-Simons number variance obtained from the
lattice simulations (points), together with the fitted curves.
Panel (b) shows the corresponding Chern-Simons diffusion rate,
obtained by differentiating the fits in panel (a).
Panel (c) shows the spatially averaged scalar background and its fitted
time evolution. Panel (d) shows the resulting effective chemical
potential defined in Eq.~\eqref{eq:chemical_potential}.
Panel (e) shows the integrand of the final fermion-number yield.
}
\label{fig:baryon-asymmetry-inputs}
\end{figure*}

\subsubsection{Generated baryon asymmetry}
We present our results in terms of the fermion number yield, estimated by using the formalism described in \SEC{sec:baryon}. 
As indicated in Eq.~\eqref{yield}, the resulting asymmetry is determined by the time integral of the product of the Chern-Simons diffusion rate and the time derivative of the spatially averaged scalar background. 
Consequently, a larger Chern-Simons variance does not necessarily lead to a larger asymmetry. 
The temporal overlap between these two quantities also plays a crucial role in determining the final yield.

The time dependence of $\Gamma_{\rm CS}(t)$ and $\dd\overline{\Phi^\dagger\Phi}(t)/\dd t$ can be extracted directly from the simulation outputs $\Delta n_{\rm CS}^2(t)$ and $\overline{\Phi^\dagger\Phi}(t)$. 
However, residual field oscillations, finite-volume effects, and finite ensemble statistics lead to late-time fluctuations in these quantities. 
Direct numerical differentiation suffers from an artificial late-time tail. 
We therefore fit the time evolution of $\Delta n_{\rm CS}^2(t)$ and $\overline{\Phi^\dagger\Phi}(t)$ with smooth saturating functions,
\begin{align}
    f(t)=f_0+\frac{\Delta f}{2} \lkk 1+\tanh\lmk \frac{t-t_0}{\tau}\rmk \rkk, \label{eq:fitting_function}
\end{align}
where $f_0$, $\Delta f$, $t_0$, and $\tau$ are determined from the simulation data. 
The fitted functions can be then differentiated analytically to obtain $\Gamma_{\rm CS}(t)$ and $\mu(t)$. 
The example cases with $\epsilon=10^{-2},10^{-3},10^{-4},\gamma_\ast=5.5$ are shown in Fig.~\ref{fig:baryon-asymmetry-inputs}.

\begin{figure}[!t]
    \centering
    \includegraphics[width=0.9\linewidth]{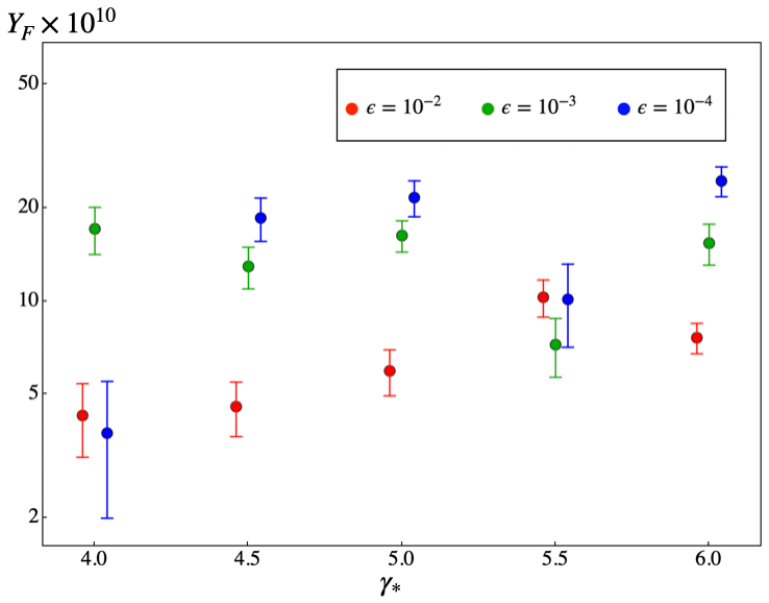}
    \caption{The result of the fermion number yield generated in the supercooled phase transition. 
    The colored points represent the central value of the estimation for different values of $\epsilon=10^{-2},10^{-3},10^{-4}$.
    Error bars denote one standard error obtained by propagating the ensemble fluctuations of the two observables $\Delta n_{\rm CS}^2$ and $\overline{\Phi^\dag\Phi}$ through the fits and the fermion number-yield integral.}
    \label{fig:baryon_yield_result}
\end{figure}

We normalize the time with the inverse of Higgs mass to show explicitly the importance of the overlapping between $\Gamma_{\rm CS}(t)$ and $\mu(t)$ in the time domain.
The difference in the ending time of simulation comes from the dependence of $R_*$ on $\epsilon$.
The final result is shown in Fig.~\ref{fig:baryon_yield_result}.
We estimate the statistical uncertainty in the fermion number yield by propagating the ensemble standard errors through the fitted observables. See Appendix~\ref{app:statistical_errors} for details.

Our estimates indicate that a sufficient asymmetry can be generated
even for $\epsilon=10^{-3}$ and $10^{-4}$ in the present benchmarks.
This differs from the small-$\epsilon$ suppression found in the
effective single-field treatment of Ref.~\cite{Bhusal:2025lvm}.
An important difference is that, in our setup, the field driving
the phase transition is separated from the Higgs field responsible
for the topological dynamics.
The non-degeneracy of the $\sigma$ potential therefore does not
necessarily lead to the same suppression of the Higgs dynamics as
in the single-field treatment.
The post-transition dynamics of $\sigma$ may further contribute to
the efficient topological production observed in our simulations.
After the transition, $\sigma$ undergoes large oscillations around
its true-vacuum value, which modulate the effective Higgs mass
through the $\sigma$-$\Phi$ coupling and can transfer energy into
the Higgs sector.
These oscillations may also induce parametric resonance and enhance
Higgs fluctuations, facilitating the formation of the Higgs
winding number and, consequently, the Chern-Simons number.
While our present simulations do not establish the occurrence or
quantitative importance of parametric resonance, the results suggest
that post-transition dynamics can play an important role in
Chern-Simons number production even for highly non-degenerate
vacua.

In addition, we find no clear monotonic dependence on $\gamma_*$ over the range explored in our simulations. This behavior may indicate that the Chern-Simons number production is not determined solely by the bubble-wall Lorentz factor at collision, but is also sensitive to the subsequent nonlinear dynamics of the Higgs-gauge system. In particular, the post-transition dynamics of $\sigma$, including the possible resonant excitation of the Higgs field, may reduce the direct sensitivity to the initial collision kinematics. Our numerical simulations are necessarily performed at values of $v_\sigma$ much larger than those expected in phenomenologically motivated supercooled phase transitions, and extrapolation to smaller $v_\sigma$ and correspondingly larger $\gamma_*$ should therefore be treated with caution. Nevertheless, the absence of a strong dependence on $\gamma_*$ within the range studied here suggests that efficient Chern-Simons number production may persist toward the more strongly supercooled regime. Together with our results for small $\epsilon$, this indicates that a strongly non-degenerate potential, which is favorable for a large GW signal, is not necessarily incompatible with efficient topological production.

\section{Discussion and Conclusion
\label{sec:Discussion}}
In the present work, we have investigated cold baryogenesis motivated by a supercooled first-order phase transition by performing three-dimensional lattice simulations of the coupled scalar-Higgs-gauge dynamics. We considered a two-field setup in which the scalar field $\sigma$ drives the phase transition, while the Higgs field $\Phi$ is coupled to $\sigma$ but does not itself control the tunneling dynamics. Starting from bubble configurations obtained from the bounce solution, we followed the subsequent bubble expansion, collisions, and nonlinear evolution of the Higgs and gauge fields, and studied the production of the Higgs winding and Chern-Simons numbers.

Our main result is that efficient Chern-Simons number production can persist even for strongly non-degenerate potentials with $\epsilon \ll 1$, which is relevant for supercooled phase transitions where a large release of vacuum energy can lead to a strong GW signal. The power spectrum of the Chern-Simons number density
shows that the characteristic scale of the topological
configurations is controlled by the Higgs-gauge sector.
We also found no clear monotonic dependence of the
Chern-Simons production on the bubble-wall Lorentz
factor $\gamma_*$ over the range accessible to our
simulations.
These observations suggest that the efficiency of
topological production is governed not only by the
initial bubble collision but also by the subsequent
nonlinear Higgs-gauge dynamics.

We have further estimated the fermion number asymmetry
by introducing a CP-violating higher-dimensional operator
perturbatively.
Our estimates show that a sufficient asymmetry can be
generated for the present benchmarks even for
$\epsilon=10^{-3}$ and $10^{-4}$.
The parameter region with strongly non-degenerate vacua
is therefore not necessarily incompatible with efficient
cold baryogenesis.

Several limitations and challenges remain in our analysis.
We have assumed $\beta/H=\mathcal{O}(10)$ rather than deriving it from the finite-temperature tunneling rate.
Moreover, the large hierarchy between the bubble separation and the microscopic wall width poses a significant challenge to directly simulating the phenomenologically motivated range of $v_\sigma$.
The CP-violating interaction is also treated perturbatively rather than included directly in the real-time lattice evolution.
A more quantitative prediction of the resulting fermion asymmetry will require a better understanding of these limitations, together with further developments in their treatments.

Another interesting direction is to study the spatial inhomogeneity of the generated fermion asymmetry. In the present analysis, we have focused on the volume-averaged asymmetry, although the underlying Chern-Simons dynamics is highly inhomogeneous, as reflected in the power spectrum of the Chern-Simons number density. Since topological transitions are induced by localized bubble collisions and the subsequent nonlinear field dynamics, the generated fermion asymmetry is expected to inherit spatial fluctuations on scales associated with the bubble dynamics. For low-scale supercooled phase transitions, such inhomogeneities may not be sufficiently erased by ordinary baryon diffusion before Big Bang Nucleosynthesis, potentially leading to additional cosmological constraints
(see e.g. Refs.~\cite{Bagherian:2025puf,Elahi:2026ldn,Girmohanta:2026ogq,Tanaka:2026ssu} for recent studies). Our lattice framework provides a direct way to determine the initial spatial distribution of the asymmetry, which can then be used to follow its subsequent transport and diffusion. A quantitative study of this connection between bubble dynamics and baryon inhomogeneity is left for future work.

Our results demonstrate that strong supercooling does not necessarily suppress the topological dynamics for cold baryogenesis.
Separating the phase-transition field from the Higgs sector allows efficient Chern-Simons number production to persist even for strongly non-degenerate vacua. This opens the possibility that a supercooled first-order phase transition can simultaneously provide a large GW signal and generate the matter-antimatter asymmetry of the Universe.

\section*{Acknowledgments}

Y.N. is supported by Natural Science Foundation of Shanghai.
N.K. is supported by JSPS KAKENHI (Grant Number 26K00695).

\appendix

\section{Lattice gauge formalism
\label{app:lattice}}

In this appendix, we summarize the lattice formulation used for the real-time simulations in Sec.~\ref{sec:simulation}. Following Ref.~\cite{Figueroa:2020rrl}, we discretize the coupled scalar-gauge system while preserving the $SU(2)$ gauge symmetry.

The gauge field is represented by the link variable,
\beq
U_\mu = \exp(-ig \delta x^\mu A_\mu) \ ,
\eeq
where we do not apply the implicit sum about $\mu$.
From the link variable, the plaquette is defined as follows,
\beq
U_{\mu\nu} \equiv U_\mu U_{\nu,+\mu} U_{\mu,+\nu}^\dagger U_\nu^\dagger \ ,
\eeq
where $U_{\nu,+\mu}\equiv U_\nu(x+\delta x\,\hat\mu)$ with $\hat\mu$ being the unit vector along $x^\mu$.
In addition, it is convenient to define
\beq
U_{\mu-\nu} \equiv U_\mu U_{\nu,+\mu-\nu}^\dagger U_{\mu,-\nu}^\dagger U_{\nu,-\nu} \ .
\eeq
The field strength tensor can be then expressed in terms of the plaquette as
\beq
F_{\mu\nu} = \frac{i(U_{\mu\nu} - U^\dagger_{\mu\nu})}{2 g \delta x^2} \ .
\eeq
The discretized lattice action of the system is now given by
\beq
\begin{split}
S &= \sum_{t,x} \delta t \delta x^3 \left[ - \frac{1}{2} \mathrm{Tr}(F_{\mu\nu} F^{\mu\nu}) - (D_\mu \Phi)^\dagger D^\mu \Phi - V(\Phi) \right] \\
&=\sum_{t,x} \frac{1}{g^2} \bigg[ \frac{\delta x}{\delta t} \sum_i \mathrm{Tr}(4-U_{0i}-U_{0i}^\dagger) \\
&\qquad \qquad - \frac{\delta t}{\delta x} \sum_{i}\sum_{j>i} \mathrm{Tr}(4 - U_{ij} - U^\dagger_{ij})\bigg] + \dots.
\end{split}
\eeq
Here we impose the temporal gauge, $A_0=0$, which reads $U_0 = I$ and then 
\beq
U_{0i} = U_{i,+0} U^\dagger_i \ . \label{eq:U0i}
\eeq
The gauge covariant derivative is discretized as
\beq
D_\mu \Phi = \frac{U_\mu \Phi_{+\mu}-\Phi}{\delta x} \ ,
\eeq
with short-hand notation $X_{\pm \mu} = X(x \pm \hat{\mu} \delta x^\mu)$ for a field variable $X$.
The evolution equations are derived from the above lattice action, which are given by
\begin{eqnarray}
F^a_{0i} 
\!\! &=& \!\! F^a_{0i,-0}  \\
&+&\!\!\! \delta t \bigg[ \frac{g}{\delta x} \mathrm{Re}(\Phi^\dagger_{+i} U^\dagger_i i\tau^a \Phi) \! - \!\frac{1}{g \delta x^3} \sum_{j \neq i} \mathrm{Tr}[i\tau^a (U_{ij} + U_{i-j})] \bigg] \nonumber
\end{eqnarray}
and
\beq
\Pi = \Pi_{-0} + \frac{\delta t}{\delta x^2} \sum_i \bigg[ U_i \Phi_{+i} + U^\dagger_{i,-i} \Phi_{-i} - 2 \Phi \bigg] \ ,
\eeq
where we define the conjugate momentum,
\beq
\Pi = \frac{1}{\delta t} (\Phi_{+0} - \Phi) \ ,
\eeq
which determines the evolution of the scalar field value,
\beq
\Phi_{+0} = \Phi + \delta t \Pi \ .
\eeq
In addition, the evolution of the link variable is determined by Eq.~\eqref{eq:U0i} as
\beq
U_{i,+0} = U_{0i}U_i \ .
\eeq
The lattice action also gives the Gauss's constraint,
\beq
\begin{split}
&\frac{1}{\delta x} \sum_i \mathrm{Tr} \big[ \tau^a (F_{0i} - U^\dagger_{i,-i} F_{0i,-i} U_{i,-i}) \big] \\
&+ g \mathrm{Re}(\Pi^\dagger i\tau^a \Phi) = 0 \ .
\end{split}
\eeq

\begin{figure}[!t]
    \centering
    \includegraphics[width=0.9\linewidth]{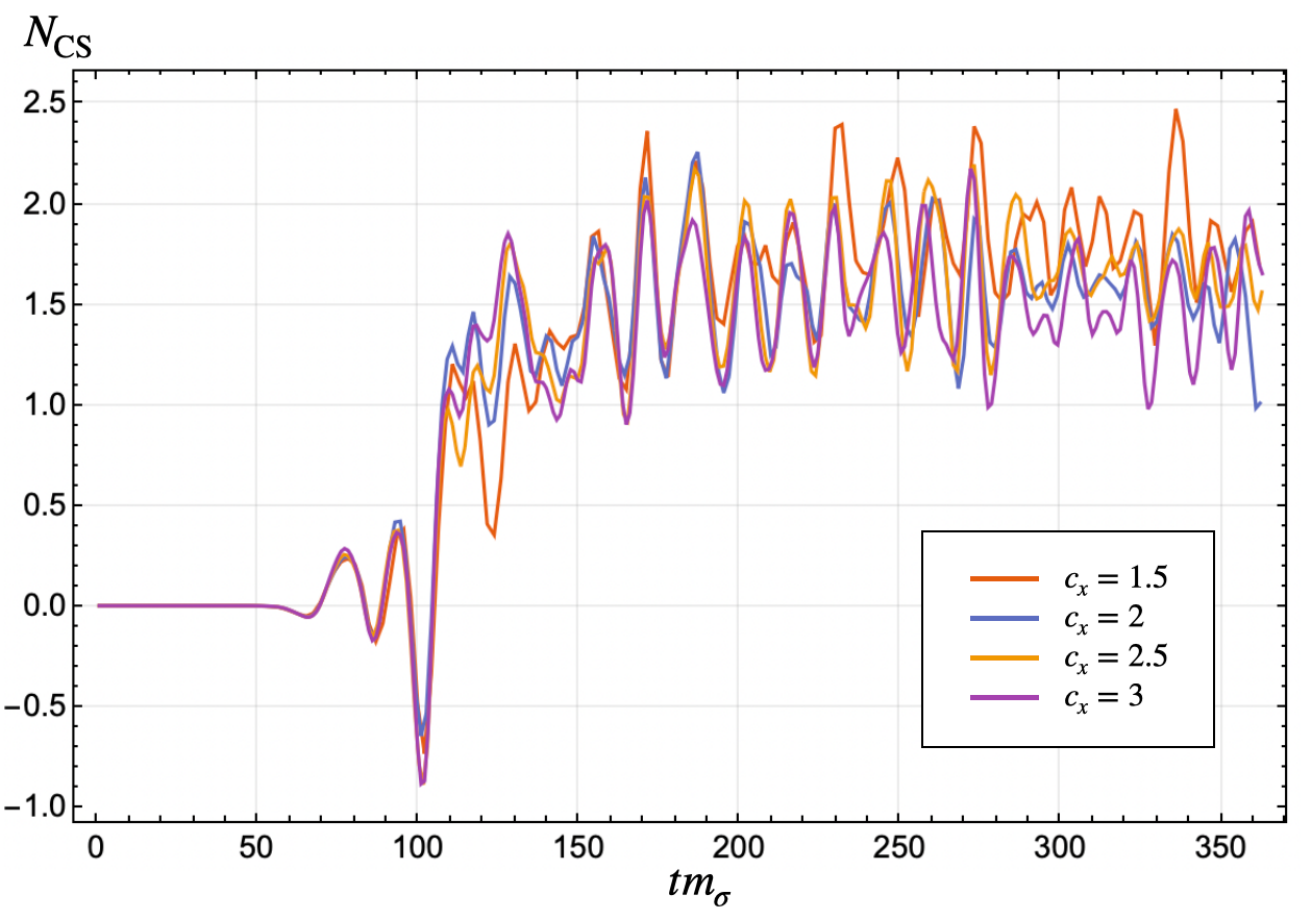}
    \caption{Time evolution of the Chern-Simons number for different spatial resolutions, characterized by $c_x=l_{\gamma_*}/\delta x$. 
    All parameters and initial conditions are held fixed. The results converge for $c_x\gtrsim2$, indicating that the production of Chern-Simons number is adequately resolved.}
    \label{fig:resolution_CS}
\end{figure}

According to Eq.~\eqref{eq:winding_number}, the lattice Higgs winding number density is defined as
\begin{align}
    n_H^{\rm lat}(x)=\frac{1}{24\pi^2}\epsilon_{ijk}\Tr[(\del_i\tilde\Phi)\tilde\Phi^\dagger(\del_j\tilde\Phi)\tilde\Phi^\dagger(\del_k\tilde\Phi)\tilde\Phi^\dagger] \, ,
\end{align}
where $\tilde\Phi$ is the discretized version of Eq.~\eqref{eq:Phi_tilde} and $\del_i\tilde\Phi$ is defined as
\begin{align}
    \frac{1}{\delta x}\lmk\tilde\Phi_{+i}-\tilde\Phi\rmk \, .
\end{align}
The total Higgs winding number is obtained by summing over the spatial lattice,
\begin{align}
    N_H^{\rm lat}(t) = \delta x^3 \sum_x n_H^{\rm lat}(t,x) \ .
\end{align}
Since the $F_{\mu\nu}\tilde F^{\mu\nu}$ term on the right-hand side of Eq.~\eqref{eq:CS_number} is a total derivative, the Chern-Simons number density can be rewritten as
\begin{align}
    n_{\rm CS} = \frac{g^2}{16\pi^2}\epsilon_{ijk}\Tr\lkk A_iF_{jk}-\frac{2i}{3}A_i A_j A_k\rkk\, .
\end{align}
Using the link and plaquette variables, the lattice Chern-Simons number density is
\begin{align}
n_{\rm CS}^{\rm lat}(x) = &\frac{1}{64\pi^2\delta x^3} \epsilon^{ijk} \Tr\bigg[ \left(U_i-U_i^\dagger\right)\times\notag\\
& \bigg(U_{jk}-U_{jk}^\dagger-\frac{1}{3}\left(U_j-U_j^\dagger\right)\left(U_k-U_k^\dagger\right)\bigg)\bigg] \ .
\end{align}
The total Chern-Simons number is given by
\beq
N_{\rm CS}^{\rm lat}(t)=\delta x^3\sum_x n_{\rm CS}^{\rm lat}(t,x) \ .
\eeq
These lattice definitions are used to evaluate the Higgs winding and Chern-Simons numbers in the simulations presented in Sec.~\ref{sec:simulation}.

\section{Systematic uncertainties
\label{app:sys_uncertanties}}

 For our lattice simulation results, the systematic uncertainties may come from: 1) finite spatial resolution $c_x$ 2) finite temporal resolution $c_t$, 3) finite-volume effect, parametrized by $L/R_*$, or equivalently, the number of bubbles $N_b$ according to Eq.~\eqref{bubble_number}.
For the temporal resolution, we adopt the ratio between the timestep and the lattice spacing $\delta t/\delta x=0.2$.
This choice provides enough precision since $\delta x=l_{\gamma_*}/c_x<m_\sigma^{-1}$ and the condition~\eqref{time_resolution} can be easily satisfied.
Given the similarity of our simulation setup to that of Ref.~\cite{Bhusal:2025lvm}, we likewise adopt $N_b=10$ bubbles, which was shown there to be sufficient for suppressing the finite-volume effects.
Therefore, we focus on the test of the spatial resolution here.

\begin{figure}[!t]
    \centering
    \includegraphics[width=0.9\linewidth]{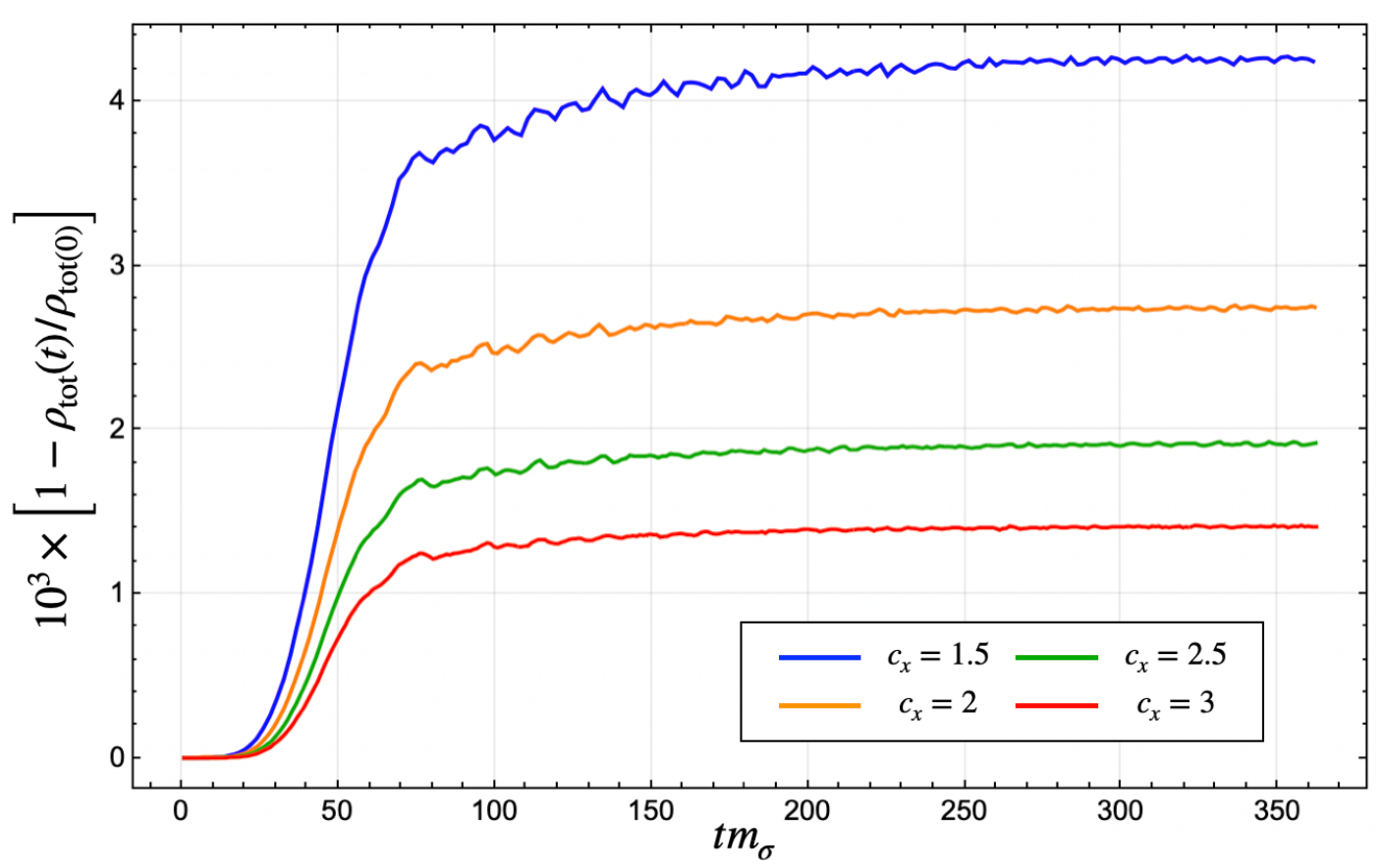}
    \caption{Relative violation of total-energy conservation for different spatial resolutions, shown as $10^3\left[1-\rho_{\rm tot}(t)/\rho_{\rm tot}^{\rm ini}\right]$. 
    The physical parameters and initial conditions are identical to those used in Fig.~\ref{fig:resolution_CS}. 
    For $c_x\gtrsim2$, the energy-conservation error remains at a controlled level throughout the evolution.}
    \label{fig:resolution_energy}
\end{figure}

As discussed in Sec.~\ref{sec:lattice_setup}, the spatial resolution in our simulation setup is characterized by the ratio between the bubble-wall width at collision and the lattice spacing $c_x=l_{\gamma_*}/\delta x$.
For given parameters, we can calculate $l_{\gamma_*}$ numerically following the discussion in Sec.~\ref{sec:bubble_expansion}.
Fixing $c_x$ then determines the lattice spacing required for the simulation.
We test different values of $c_x$ by inspecting the generated Chern-Simons number and the violation of energy conservation in the same simulation setting (parameters and initial conditions).
Figs.~\ref{fig:resolution_CS},~\ref{fig:resolution_energy} show their respective time evolution.
The results indicate that $c_x\gtrsim2$ is sufficient to resolve Chern-Simons number production reliably while maintaining the energy-conservation error at a controlled level.

\section{Statistical uncertainty}
\label{app:statistical_errors}
We estimate the statistical uncertainty of $Y_F$ directly from the ensemble of independent simulations.
The uncertainty is propagated through the fitted observables using a first-order approximation.

For $\mathcal{N}$ simulations evaluated at common times $t_i$, we define $X_{r,i}=N_{\rm CS}^{(r)}(t_i)$ and $H_{r,i}=\overline{\Phi^\dagger\Phi}^{\,(r)}(t_i)$, where $1\leq r\leq \mathcal{N}$ labels the simulation and the overline denotes a spatial average. 
The ensemble observables entering the fits are
\begin{align}
    \bar X_i = \frac{1}{\mathcal{N}}\sum_{r=1}^{\mathcal{N}}X_{r,i}\, , V_i &= \frac{1}{\mathcal{N}(Lm_h)^3}\sum_{r=1}^{\mathcal{N}}(X_{r,i}-\bar X_i)^2, \notag\\
    \bar H_i &= \frac{1}{\mathcal{N}}\sum_{r=1}^{\mathcal{N}}H_{r,i}.
\end{align}
Here, $V_i$ is the dimensionless Chern-Simons variance density.
The scalar observable is expressed in the simulation's field units, with the physical normalization restored in the yield calculation.

The fluctuations associated with each run are characterized by
\begin{align}
    u^V_{r,i} &=
    \frac{(X_{r,i}-\bar X_i)^2}{(Lm_h)^3}-V_i,
    &
    u^H_{r,i} &= H_{r,i}-\bar H_i.
\end{align}
The first expression is the estimated first-order influence of a run on the Chern-Simons variance, rather than on the mean Chern-Simons number.
Combining all $\mathcal{M}$ time points into $\mathbf u_r=(u^V_{r,1},\ldots,u^V_{r,M}, u^H_{r,1},\ldots,u^H_{r,M})^T$, we estimate the covariance of the ensemble observables as
\begin{align}
    C_y \simeq \frac{1}{\mathcal{N}(\mathcal{N}-1)}
    \sum_{r=1}^{\mathcal{N}}\mathbf u_r\mathbf u_r^T.
    \label{eq:observable_covariance}
\end{align}

Let $\mathbf p$ collect the parameters of the two unweighted fits~\eqref{eq:fitting_function}, and let $J_{ij}=\partial f_i/\partial p_j$ be the Jacobian of their combined predictions, evaluated at the best fit.
In the leading-order approximation,~\cite{Bruno:2022mfy}
\begin{align}
    \delta\mathbf p &\simeq J^+\delta\mathbf y,
    &
    C_p &\simeq J^+ C_y (J^+)^T,
    \label{eq:parameter_covariance}
\end{align}
where $J^+$ is the Moore-Penrose pseudoinverse,
\begin{align}
    J^+\equiv (J^T J)^{-1} J^T\, .
\end{align}
Writing the baryon-yield estimate as $Y_F=F(\mathbf p)$, its propagated variance is
\begin{align}
    \sigma_{Y_F}^2 \simeq
    (\nabla_{\mathbf p}F)^T C_p\,\nabla_{\mathbf p}F,
    \label{eq:yield_statistical_error}
\end{align}
with the gradient evaluated using central finite differences.
The plotted error bars in Fig.~\ref{fig:baryon_yield_result} represent $\sigma_{Y_F}$.
They describe the sampling uncertainty propagated through the chosen fitting procedure. 
Bias from the fitting ansatz, lattice artifacts, and uncertainties in the physical input parameters are not included.

\bibliography{reference}

\end{document}